\documentstyle[12pt,epsfig]{article}

\newskip\humongous \humongous=0pt plus 1000pt minus 1000pt
  \newif\ifdtup

\def\frac#1#2{ {{#1} \over {#2} }}

\def\ie{\hbox{\em i.e. }}

\def\beq{\begin{equation}}
\def\eeq{\end{equation}}

\def\beqn{\begin{eqnarray}}
\def\eeqn{\end{eqnarray}}

\def\a0{\alpha_{\sf 0}}

\begin{document}
\begin{titlepage}
\par \vskip 10mm
\begin{center}
{\Large \bf
High-loop perturbative renormalization constants \\
for Lattice QCD (I): \\
finite constants for Wilson quark currents.}
\end{center}
\par \vskip 6mm
\begin{center}
F.\ Di Renzo$\,^a$,
V.\ Miccio$\,^b$, C.\ Torrero$\,^c$ 
and  L.\ Scorzato$\,^{d}$ \\
\vskip 10 mm
$^a\,${\it Dipartimento di Fisica, Universit\`a di Parma \\
and INFN, Gruppo Collegato di Parma, Italy}\\[.5 em]
\vskip 2 mm
$^b\,${\it I.N.F.N., Sezione di Milano Bicocca, Italy}
\vskip 2 mm
$^c\,${\it Faculty of Physics, University of Bielefeld, Germany}
\vskip 2 mm
$^d\,${\sl ECT*, Villazzano (Trento), Italy} \\[.5 em]
\end{center}
\par \vskip 2mm
\begin{center} {\large \bf Abstract} \end{center}
\begin{quote}
We present a high order perturbative computation of the renormalization constants $Z_V$, $Z_A$ and of the ratio $Z_P/Z_S$ 
for Wilson fermions.  
The computational setup is the one provided by the \emph{RI'-MOM} scheme. Three- and four-loop expansions are made 
possible by Numerical Stochastic Perturbation Theory. 
Results are given for various numbers of flavors and/or (within a finite accuracy) for generic $n_f$ up to three loops. For the case $n_f=2$ we also present four-loop results. 
Finite size effects are well under control and the continuum limit is taken by means of 
\emph{hypercubic symmetric Taylor expansions}. The main indetermination comes from truncation errors, 
which should be assessed in connection with convergence properties of the series. The latter is best discussed 
in the framework of Boosted Perturbation Theory, whose impact we try to assess carefully. Final results and their  uncertainties show that high-loop perturbative computations of Lattice QCD Renormalization Constants (RC's) are feasible and should not be viewed as a second choice. 
As a by-product, we discuss the perturbative expansion for the critical mass, also for which results are 
for generic $n_f$ up to three loops, while a four-loop result is obtained for $n_f=2$.
\end{quote}

\end{titlepage}

\section{Introduction}
Lattice Perturbation Theory (LPT) has been for a long time the only available tool for the computation 
of Lattice QCD Renormalization Constants (RC's). By now, non-perturbative computations are preferred. 
We should stress, however, that there is no theoretical obstacle to the perturbative computation of 
either \emph{finite} or \emph{logarithmically divergent} RC's like (for example, those for quark 
bilinears or their ratios). The main difficulties are of practical nature. The first one is that LPT is 
technically very hard, much harder than Perturbation Theory (PT) on the continuum \cite{StefC}. 
Therefore, computations are often performed only at one loop. This is a serious limitation, which 
is made even more severe by the bad convergence properties of LPT. 
To take care of this problem Boosted Perturbation Theory (BPT)  and/or of Tadpole-Improved Perturbation 
Theory (TIPT) \cite{BPT} is often used. There is quite a consensus on the fact that, at one loop 
the impact of BPT (and/or TIPT) is often important. On the other side, there is no clear-cut result 
on the actual control on these procedures. One should always keep in mind that convergence properties 
of the series are the real issue and assessing them from a one-loop computation is of course impossible.
Other improvement schemes have been in recent years proposed, which aim at resumming some leading 
contributions \cite{HarisCactus}.
A different approach to the computation of RC's in Lattice QCD is a completely non-perturbative one. 
In this case, one needs an intermediate scheme, which is eventually matched to the $\overline{\rm MS}$ 
scheme (the one in which phenomenologists are most interested in) by a continuum perturbative 
computation. Popular intermediate schemes are the Regularization Independent (\emph{RI'-MOM}) 
\cite{RI-MOM-rome} scheme and the Schr\"odinger Functional (\emph{SF}) \cite{SF-alpha} scheme.
A non-perturbative computation eliminates the truncation errors. On the other side, one needs to 
face all the difficulties inherent in numerics. Among these, the high computational effort
of unquenched (or partially quenched) Lattice QCD simulations. 
In practice it is sometimes extremely hard to get a good signal for realistic simulation parameters.
For this reason LPT is still necessary, either for comparison or because the only feasible approach.

In recent years the technique of Numerical Stochastic Perturbation Theory (NSPT) has been introduced 
(for an extended introduction - which in particular covers the unquenched version - see 
\cite{NSPTfull}). NSPT is a numerical implementation of Stochastic PT \cite{ParWu}. It is a numerical 
tool which enables to perform LPT computations with no reference whatsoever to diagrammatics. 
By making use of NSPT we can compute Lattice QCD RC's to high orders, which here means 3 (or even 4) 
loops. At these orders the use of BPT enables to assess the convergence properties of the series and 
better control the truncation errors. Since we necessarily work on a finite lattice, finite-volume 
and scaling violation effects have to be assessed carefully: this can be done. A careful extraction of 
continuum limit is one of the good point of the approach: the solution comes from what we call 
\emph{hypercubic symmetric Taylor expansions}. Another nice feature comes from the fact that we
can work directly in the massless limit (which is also where RC's are usually defined).
This also eliminates the need of an expensive chiral extrapolations. 
Finally, perturbative computations offer the possibility of a stronger analytical control, 
knowing the dependence on the coupling and on the number of flavors.  

The main message of this paper is that high-loop perturbative computations of Lattice QCD RC's are 
feasible and should not be seen as a second choice. In particular, having both the perturbative and 
the non-perturbative determinations of a RC's gives a valuable comparison. This is not at all academic.
As a matter of fact, non-perturbative determinations are based on assumptions that are only proved in 
PT. 

This is the first of a couple of papers which deal with the NSPT perturbative computation of Wilson 
quark bilinears. Renormalization conditions are fixed by the \emph{RI'-MOM} prescriptions. 
Here we can make comparison with a non-perturbative determination \cite{CeciVitt}.\footnote{
The comparison will be made for given values of the coupling ($\beta=5.8$) and number of flavors 
($n_f=2$).} In this paper we will concentrate on the determination of \emph{finite} RC's: $Z_V$, $Z_A$ 
and the ratios $Z_P/Z_S$ and $Z_V/Z_A$ \footnote{We will explain below why the computation of $Z_V$, 
$Z_A$ and $Z_V/Z_A$ are not tautological here.} for (unimproved) Wilson fermions. As a by-product, 
we also obtain the expansion for the critical mass. Results are given for various numbers of flavors. 
At three loops some results are even given (to a finite accuracy) for generic $n_f$. Instead, we 
present fourth loop results for the $n_f=2$ case only. A different paper will deal with the computation
of logarithmically divergent RC's for quark bilinears (in particular, the RC for the scalar current 
$Z_S = Z_m^{-1}$, which is phenomenologically relevant for the determination of the quark masses). 
This deserves some extra caution, since dealing with anomalous dimensions requires a peculiar care for 
finite volume effects. 

The paper is organized as follows. In section 2 we recall the basic definitions of the renormalization 
scheme to which we adhere, while in section 3 we discuss some technical details of our computations. 
Section 4 introduces the main tool which is needed to extract the continuum limit (the already 
mentioned \emph{hypercubic symmetric Taylor expansions}): this is done by discussing the (prototype) 
computation of the quark propagator. Section 5 contains our results: first we discuss the finite ratios 
$Z_P/Z_S$ and $Z_V/Z_A$, for which we can fit three loops results for generic $n_f$; then we move to 
$Z_V$ and $Z_A$ (results are given at three loops for $n_f = 0$ and at four loops for $n_f=2$); 
finally we present a by-product of our computations, \emph{i.e.} the critical mass to three loops 
(again, actually four in the case $n_f=2$). In section 6 we discuss the general features of 
computations dealing with an anomalous dimension (this sets the stage for  what will be discussed in a 
following paper \cite{ourNEXT}). In section 7 we deal with resummations and convergence properties of 
our series and finally section 8 contains our conclusions and perspectives for future applications. 

\section{The RI'-MOM renormalization scheme}

In order to compute renormalization constants we adhere to the \emph{RI'-MOM} scheme. This is one of 
the so-called  physical schemes\footnote{One should nevertheless keep in mind that the name 
\emph{physical} is actually misleading in the case of QCD.} (as opposed to the more popular 
$\overline{\rm MS}$ scheme) and goes back to the \emph{MOM} scheme of \cite{CG79}. It became very 
popular after the introduction of non-perturbative renormalization in \cite{RI-MOM-rome}. 
\emph{RI} emphasizes the \emph{regulator independent} nature of the scheme, which in particular makes 
the lattice a viable regulator. The prime denotes a renormalization condition for the quark field 
which is slightly different from the original one. All the details on this scheme can be found, 
for example, in \cite{JGrac}. In the following we only introduce the definitions which are relevant 
for our application. 

The basic quantities of our computation are of quark bilinears between external quark states at 
fixed (off-shell) momentum $p$
\begin{equation}\label{basicC}
\int dx \,\langle p | \; \overline{\psi}(x) \Gamma \psi(x) \; | p \rangle \, = \, G_{\Gamma}(pa). 
\end{equation}
Here $\Gamma$ stands for any of the $16$ matrices, which provide the standard basis of the Dirac 
space (Dirac ind ices will often be suppressed). We adopt the usual naming convention for the 
bilinear: the scalar ($S$) is defined by $\Gamma=1$, the vector ($V$) by $\Gamma=\gamma_\mu$, 
the pseudoscalar ($P$) by $\Gamma=\gamma_5$,  the axial ($A$) by $\Gamma=\gamma_\mu \gamma_5$ 
and the tensor ($T$) by $\Gamma=\sigma_{\mu\nu}=1/2\; [\gamma_\mu, \gamma_\nu]$.
Above, we made explicit the dependence on the lattice spacing $a$, which serves as a regulator. 
Later we will use the notation $\hat{p}=pa$.

\noindent Being these quantities gauge-dependent, a choice for the gauge condition has to be made. We will focus on computations in the Landau gauge. From a numerical point of view, this gauge condition is easy to fix on the lattice. On top of that, one does not need to discuss the gauge parameter renormalization. It also gives some extra bonus: the anomalous dimension for the quark field is zero at one loop.  

\noindent We can trade the $G_{\Gamma}(pa)$ for the amputated function $\Gamma_{\Gamma}(pa)$ 
($S(pa)$ is the quark propagator)
\begin{equation}
G_{\Gamma}(pa) \;\;
\rightarrow \;\; \Gamma_{\Gamma}(pa) \, = \, S^{-1}(pa) \; G_{\Gamma}(pa) \; S^{-1}(pa).
\end{equation}
The $\Gamma_{\Gamma}(pa)$ are eventually projected on the tree-level structure by a suitable operator $\hat{P}_{O_{\Gamma}}$
\begin{equation}
	O_{\Gamma}(pa) = \mbox{Tr}\left(\hat{P}_{O_{\Gamma}} \; \Gamma_{\Gamma}(pa)\right).
\end{equation}
Renormalization conditions are now given in terms of the $O_{\Gamma}(pa)$ according to 
\begin{equation}\label{master}
	Z_{O_{\Gamma}}(\mu a, g(a)) \, \; Z_q^{-1}(\mu a, g(a)) \, \; O_{\Gamma}(pa) \Big|_{p^2 = \mu^2} \, = \, 1.	
\end{equation}
Here the $Z$'s depend on the scale $\mu$ via the dimensionless quantity $\mu a$, while the 
dependence on $g(a)$ will be expanded in PT. 
One should keep in mind from the very beginning that we will be eventually interested in the 
$a \rightarrow 0$ limit of the $Z$'s.
The quark field renormalization constant $Z_q$, which enters the above formula, is defined by
\begin{equation}\label{Zq}
	Z_q(\mu a, g(a)) \, = \, - i \frac{1}{12} \frac{\mbox{Tr}(\hspace{-.2em}\not\hspace{-.2em}p \; S^{-1}(pa))}{p^2}\Big|_{p^2 = \mu^2}.
\end{equation}
The original \emph{RI} scheme (without a prime), would have a derivative with respect to $p^{\mu}$,
instead. 

\noindent In order to get a mass-independent renormalization scheme, one imposes renormalization conditions on massless quarks. In Perturbation Theory this implies the knowledge of the relevant counterterms, \ie the values of the various orders of the Wilson fermions critical mass. One- and two-loop results are known from the literature \cite{PanaPelo}. Third (and fourth) loop have been computed by us as a (necessary) by-product of the current computations: results are reported in section 5 (the three loop result in the $n_f=2$ case has already been reported in \cite{NSPTfull}). Notice that the situation in the non-perturbative framework is more cumbersome with respect to staying in the massless limit. The determination of the critical mass is in a sense the prototype non-perturbative computation of a(n additive) renormalization constant. As it is well known, this is also a matter of principle: being a power-divergent renormalization, the critical mass itself can not be computed in Perturbation Theory in the continuum limit. Still, from a numerical point of view the massless limit is always reached by an extrapolation procedure, which is usually a major source of error in non-perturbative determinations of RC's for Lattice QCD. 

A great advantage of working in the \emph{RI'-MOM} scheme is that the relevant anomalous dimensions are known to 3 loops \cite{JGrac}. One is usually ready to admit that getting the logarithms is the \emph{easy} part in the computation of a renormalization constant, while fixing the finite parts is the \emph{hard} part of the work. As we will see, the situation is, to a certain extent, the opposite in the case of NSPT. We actually take for granted the logarithms (they are fixed by the choice of the scheme) and mainly concentrate on the computation of finite parts. As it is discussed in section 6, finite size effects open anyway the backdoor to corrections to the logarithmic contributions. Being 3 loops the order to which anomalous dimensions are known, this is also the order at which we can push our computations for every observable which has a non-vanishing anomalous dimension. On the other side, the finite RC's we will be concerned with in the present paper are in principle not constrained by anything but numerical precision, and that is why we pushed the computation of these quantities to an even higher order (4 loops, at the moment).  

\section{Some technical details of our computations}

The lattice formulation we use in this work is defined by the plain Wilson action for gauge fields and plain 
(\emph{i.e.} unimproved) Wilson fermions. 
As said, our computational tool is NSPT \cite{NSPTfull}. Here we only point out those technical details
which are relevant to the present computation. In its actual implementation, NSPT shares a few 
ingredients which are common to any lattice simulation. The main peculiarity is the representation of 
the fields as an expansion in the coupling constant, \emph{i.e.} 
\begin{equation}
	U_\mu(x) = 
  1 + \sum_{i=1}^{n} \, \beta^{-\frac{i}{2}} \; U_\mu^{(i)}(x) 
  \;. 
\end{equation}
As it is apparent from the formula above, our preferred expansion parameter is the inverse of the lattice parameter $\beta = 2N_C/g_0^2$, $N_C$ being the number of colors and $g_0=g(a)$ the bare lattice coupling; thus $\beta^{-\frac{1}{2}}$ is  proportional to $g_0$.
In our case a three-loop computation requires $n=6$, while for four loops (which is at the moment the maximum order for which we report results in the $n_f=2$ case) one needs $n=8$. 
The proliferation of fields results in the request of a bigger amount of memory than in ordinary (non-perturbative) lattice QCD simulations. It is of course relevant also in terms of computing power: the algorithm is dominated by order-by-order multiplications, \emph{i.e.} the number of floating-point operations grows as $n(n-1)/2$. While this could appear as a big overhead with respect to ordinary non-perturbative dynamics, this is actually not true. In particular, in unquenched NSPT (like in any fermionic simulation) the basic building block is the inversion of the Dirac matrix, for which the perturbative nature of the computation results in a \emph{closed} recursive algorithm, which is fairly well implemented (see \cite{NSPTfull}). On top of that, as we will discuss a bit more later, there is no need for an extrapolation to the chiral limit. As a result, NSPT fermionic computations are actually less demanding than non-perturbative counterparts. 

As in many non-perturbative numerical computations, it is worth producing fairly decorrelated configurations and store them for different subsequent measurements. A $32^4$ lattice (both at three and at four loops) fits well on an \emph{APEmille} crate. At the same orders, a $16^4$ lattice can be managed by small \emph{PC}-clusters or even by a robust (but nowadays standard) \emph{PC}. While the first case is treated by our \emph{TAO}\footnote{\emph{TAO} is the \emph{APE}-dedicated programming language.} codes, the second is implemented in the framework of a by now well-established $C^{++}$ NSPT package.  

The number of flavors $n_f$ enters the computations as a parameter, \emph{i.e.} one has to perform different simulations for different $n_f$'s. In Perturbation Theory each order has a trivial polynomial dependence on $n_f$, so that one can fit the $n_f$ dependence. $n_f=0$ has by now been simulated both on $16^4$ and on $32^4$ lattices: results have been used to assess finite-size effects. The unquenched cases have been simulated on the bigger ($32^4$) lattice: $n_f=2$ is the case for which we have the largest number of configurations, while we also have several tenths of configurations for both $n_f=3$ and $n_f=4$. As a result, at the moment we are not going to quote every result for any $n_f$. In particular, four-loop results are at the moment only given in the case $n_f=2$, for a reason that will be clear in a moment.

As already stated, one good feature of NSPT computations is the fact that one can stay at the chiral limit. As we have already pointed out, the computation of the Wilson fermion critical mass was in a sense the prototype computation of a non-perturbative (additive, in this case) renormalization constant. It is also the prototype of a power-divergent renormalization, which can not be safely computed in PT in the continuum limit. On the other hand, no numerical simulation can be performed at $k_{critical}$ (we adhere to the common non-perturbative notation of quoting the hopping parameter rather than the mass of the quark): the chiral limit is always reached by means of a convenient chiral extrapolation. \\

In Perturbation Theory one corrects for the additive quark mass renormalization by plugging in critical mass counterterms order by order. This is exactly what we do in NSPT. We were ready to start our simulations straightaway at three-loop order, which requires the knowledge of the critical mass up to two loops, and this is exactly what can be taken from the literature \cite{PanaPelo}. Each subsequent order asks for an iterative procedure: one computes the critical mass at the $n^{th}$ order (from $n^{th}$-order simulations) and then plug it in the $(n+1)^{th}$-order simulations. In particular, for the case $n_f=2$ our determination of the three-loop critical mass was good enough to plug it into four-loop simulations. The statistics we collected for the other values of $n_f$ are at the moment not sufficient to safely aim at the same accuracy. 

The \emph{RI'-MOM} scheme renormalization has been discussed in a generic covariant gauge \cite{JGrac}. We have already stated that our computations were performed in Landau gauge and stressed what the advantages of a such a choice are. From the point of view of computer simulations fixing the gauge  to Landau in NSPT simply requires the order-by-order implementation of a well-known ($FFT$-accelerated) iterative procedure (for details, see \cite{NSPTfull}). It is worth stressing that 
in the NSPT framework also a peculiar implementation of the Faddeev-Popov mechanism is possible (see \cite{3d} for an application): by the same trick which enables us to treat the fermionic determinant we can manage the Faddeev-Popov determinant, without the inclusion of ghost fields. Still, we can compute in any covariant gauge with gauge parameter $\xi\neq0$, \emph{i.e.} Landau gauge is the only one which is not viable (apart from an extrapolation procedure). While the generic covariant gauge NSPT simulation has a (moderate) computational overhead, Landau gauge-fixing has a delicate issue in the numerical noise which is introduced by the (order-by-order) iteration. We explicitly checked that this noise was not a big problem (of course $FFT$-acceleration is quite helpful in reducing the number of iterations needed to fix the gauge). In the end the advantages of computing in Landau gauge were not overtaken by the care that is due to keep this noise under control.

We now come to briefly describe how we compute the observables of Eq.~(\ref{basicC}). 
Trivial algebra (\emph{i.e.} creating external states with quark operators and Wick-contracting to obtain propagators) leaves us with the task of computing expectation values (\emph{i.e.} asymptotic Langevin time averages) of the quantities 
\begin{equation}
\sum_{q;\sigma \tau} M^{-1}_{\alpha p; \sigma q} \; 
\Gamma_{\sigma \tau} \; M^{-1}_{\tau q; \beta p} .
\end{equation}
(where $M$ is the Dirac operator; $\alpha$ and $\beta$ are external polarizations; $\sigma$ and $\tau$ are 
other spin ind ices; $p$ and $q$ are momentum indices; color degrees of freedom are always suppressed in the 
notation)
The index $p$ in the inverse Dirac operator is singled out by placing a $\delta$-like source at $p$ in momentum space, with the right polarization and color index (more details in the following section). Notice that in this way not only the inverse is to be computed on a source (as usual), but one actually squeezes all the information out of the configuration. This is the advantage of working directly in momentum space, which is natural in our framework (every inversion of $M$ comes as a result of a computation which goes back and forth from momentum space; again, see \cite{NSPTfull} for details). The only measurement which is a bit different is that of the conserved vector current
\begin{equation}
	V_\mu^c = 1/2 \; \left( \, \overline{\psi}(x) \,(\gamma_\mu-1)\, U_\mu(x)\, \psi(x+\mu) \, + \, 
	\overline{\psi}(x+\mu) \,(\gamma_\mu+1)\, U_\mu^{\dag}(x)\, \psi(x) \,\right) \, .
\end{equation}
A little algebra shows that also in this case the measurement can be quite efficient by reverting to a convolution product.

A very important improvement of our statistics comes from exploiting hypercubic symmetry: all the measurements connected by a hypercubic symmetry transformation are averaged. The fluctuations associated to this average are taken into account for assessing errors. As a general rule for the different measurements involved in our calculations, bootstrap was the basic tool for the computation of errors.

To conclude this section about our computational method, we should comment on our treatment of the
zero modes. Any perturbative expansion of LQCD has to face the problem of regularizing the zero 
modes contribution to the functional integral, since the free propagator cannot be inverted in 
those points. How this applies to NSPT has been discussed in \cite{NSPTfull}, where we refer the 
reader for further details. The most common approach is to remove the degrees of freedom associated
with the zero modes \cite{Heller:1984hx,StefC}. Although this prescription is not gauge invariant, 
such contributions are expected to vanish in the infinite volume limit. 
We should remark that gauge invariant alternatives to this procedure exist. They involve the
use of twisted boundary conditions \cite{TwBC} or the Schr\"odinger Functional scheme \cite{SF-alpha}.
While we plan to perform computations also in those schemes in the future, in the present
work we have only considered the prescription in which zero modes are removed.
We will come back to this issue in section \ref{sec:anomdim}.

\section{Hypercubic symmetric Taylor expansions: the case of the quark propagator}

We now proceed to discuss in detail a prototype computation, \emph{i.e.} the one loop computation 
of the quark field renormalization constant. In practice, we are going to describe how we measure 
the quark propagator. 
We will thus make clear what we mean by \emph{hypercubic symmetric Taylor expansions}. 

The section is intended as a prototype computation, so let us pin down in general what the expected form of the $n^{th}$ loop coefficient of a RC is: 
\begin{equation} \label{CC}
	z_n = c_n + \sum_{i=1}^{n} d_i(\gamma) \log(\hat{p})^i + F(\hat{p}) \;\;\;\;\;\;\;
\left(\hat{p} = p a \right).
\end{equation}
We have to look for a finite number ($c_n$), a divergent part which is a function of anomalous dimensions $\gamma$'s and irrelevant pieces, which we can expect compliant to hypercubic symmetry and described by a suitable function $F$.
We take the needed anomalous dimensions from the literature and we subtract their contribution.
In particular, for a one-loop computation we simply need to subtract a simple $log$ multiplied by 
the one-loop anomalous dimension (in this section we will completely ignore all the contributions 
coming from finite-size effects, to which we will come back in Section 6). 
After such a subtraction we need a convenient way to fit the irrelevant pieces given by  
$F(\hat{p})$. The example at hand is both instructive and simple: in particular, in Landau gauge 
the quark field has zero anomalous dimension at one loop, so it is simply required that we get 
rid of $F(\hat{p})$ in order to get the constant $c_n$ we are interested in.

We want to compute the two points vertex function (the inverse of the quark propagator) 
for a massless fermion. In the continuum limit we have:
\[
\Gamma_2(p^2) = S(p^2)^{-1}.
\]
On the lattice we define the dimensionless quantity $\hat{p}=pa$ (in general we use the {\em hat}
notation for dimensionless quantities).
Furthermore, we also explicitly write the dependence on the coupling 
(and since we compute in PT we write $\beta^{-1}$ rather than $\beta$)
\begin{eqnarray}
a \Gamma_2(\hat{p},\hat{m}_{cr},\beta^{-1}) &=& a S(\hat{p},\hat{m}_{cr},\beta^{-1})^{-1} 
\nonumber \\
&=& i\hspace{-.2em}\not\hspace{-.2em}\hat{p} + \hat{m}_W(\hat{p}) - \hat{\Sigma}(\hat{p},\hat{m}_{cr},\beta^{-1}) 
\label{Prop2Self}
\end{eqnarray}
where $\hat{m}_W(\hat{p}) = {\cal O}(\hat{p}^2)$ is the (irrelevant) mass term generated at 
tree level by the Wilson prescription,  $\hat{\Sigma}(\hat{p},\hat{m}_{cr},\beta^{-1})$ is the 
dimensionless self-energy (which is ${\cal O}(\beta^{-1})$) and $\hat{m}_{cr}=a m_{cr}$ is the critical mass 
(which is ${\cal O}(\beta^{-1})$ as well).
Since chiral symmetry is broken by the Wilson regularization, also massless fermions generate
a mass counterterm.

\vspace{0.5cm}

\noindent The first step is to compute the self energy $\hat{\Sigma}$ from our
NSPT simulations. For that we need the propagator 
$a S(\hat{p},\hat{m}_{cr},\beta^{-1})$ in momentum space, \emph{i.e.}
\[
a S(\hat{p},\hat{m}_{cr},\beta^{-1})_{\alpha\eta} = 
\langle \, M^{-1}_{\alpha p;\eta p} \,\rangle = 
T^{-1} \sum_{t=1}^T \;  M^{-1}_{\alpha p; \eta p}(t) \, ,
\]
where $M_{\alpha p;\eta q}$ is the full fermionic matrix. We write explicitly only the
spin indices ($\alpha$ and $\eta$) and the momentum coordinates ($p$ and $q$), while
color indices are left implicit.
The symbols $\langle \rangle$ stand for the average over the gauge configurations
and the right hand side makes explicit the average over the Monte Carlo history 
of length $T$. This is performed as described in \cite{NSPTfull}. Here we only
remind that our method -- based on a discretized stochastic Langevin equation --
also involves an extrapolation on the stochastic time discretization. 
We are interested in those elements of the inverse fermionic matrix which appear in the main 
diagonal in momentum space. 
This is obtained by ``sandwiching'' the fermionic matrix in a $\delta$-like source vector in momentum 
space: $\xi^{(\alpha;p)}_{\sigma}(q) = \delta_{\alpha\sigma} \delta_{pq}$.
The order-by-order inversion is then performed as described in \cite{NSPTfull}.

Once $a S(\hat{p},\hat{m}_{cr},\beta^{-1})$ is obtained, we average over 
all the components which are connected by hypercubic symmetry transformations.
For each given momentum, we numerically (order-by-order) invert the $4 \times 4$ 
propagators\footnote{Here, we use the fact that the propagator is color diagonal at any order in 
Perturbation Theory}.
Finally we obtain the self energy $\hat{\Sigma}(\hat{p},\hat{m}_{cr},\beta^{-1})$ as in 
Eq.~(\ref{Prop2Self}).

\vspace{0.5cm}

Now we turn to the analysis of the self energy that we have obtained as above.
It can be written as
\begin{eqnarray}
\label{SelfE}
\hat{\Sigma}(\hat{p},\hat{m}_{cr},\beta^{-1}) = \hat{\Sigma}_c(\hat{p},\hat{m}_{cr},\beta^{-1}) + 
\hat{\Sigma}_V(\hat{p},\hat{m}_{cr},\beta^{-1}) +
\hat{\Sigma}_{\rm other}(\hat{p},\hat{m}_{cr},\beta^{-1}).
\end{eqnarray} 
$\hat{\Sigma}_c$ is the contribution along the (Dirac) identity operator. By this we mean that
the trace over spin indices: $1/4 {\rm Tr_{spin}}(\hat{\Sigma}) = \hat{\Sigma}_c$.
Similarly, $\hat{\Sigma}_V$ is the contribution along the gamma matrices:
\[
\frac{1}{4} \sum_{\mu} \gamma_{\mu} {\rm Tr_{spin}}( \gamma_{\mu} \hat{\Sigma} ) = \hat{\Sigma}_V
\]
Finally, $\hat{\Sigma}_{\rm other}$ includes all contributions along the remaining elements of
the Dirac basis. We are not interested in such (irrelevant) terms, which are easily projected out.
Therefore, we will forget about $\hat{\Sigma}_{\rm other}$ in the following.

$\hat{\Sigma}_c$ contains the contribution to the critical mass, in fact:
\begin{equation}\label{Mc}
	\hat{\Sigma}(0,\hat{m}_{cr},\beta^{-1}) = \hat{\Sigma}_c(0,\hat{m}_{cr},\beta^{-1}) = \, \hat{m}_{cr} = a \,m_{cr}.
\end{equation}
By restoring physical dimensions one can inspect the $a^{-1}$ divergence of the critical mass: $a^{-1} \, \hat{\Sigma}_c(0,\hat{m}_{cr},\beta^{-1}) = m_{cr}$. We will come back to it in the following section. 
For the moment, we concentrate on $\hat{\Sigma}_V$, which we need to extract 
the quark field RC. 
If we make a Taylor expansion in powers of $a$, its most general form up to order $O(a^4)$ is:
\begin{equation}\label{hypTayl}
\hat{\Sigma}_V=
i \sum_{\mu} \gamma_{\mu} \, \hat{p}_{\mu} \; \left( \hat{\Sigma}_V^{(0)}(\hat{p},\hat{m}_{cr},\beta^{-1}) + \hat{p}_{\mu}^2 \,\hat{\Sigma}_V^{(1)}(\hat{p},\hat{m}_{cr},\beta^{-1}) 
+ \hat{p}_{\mu}^4 \,\hat{\Sigma}_V^{(2)}(\hat{p},\hat{m}_{cr},\beta^{-1}) + \ldots \right).
\end{equation}
where the dots stand for higher terms in $a$. 
The functions $\hat{\Sigma}_V^{(i)}(.)$ (with $i=0,...$) in turns, are the most
general combinations of hypercubic-invariant polynomia which contribute to the given order.
In particular the first term can be written as:
\begin{equation}
\hat{\Sigma}_V^{(0)}(\hat{p},\hat{m}_{cr},\beta^{-1}) = 
\alpha^{(0)}_1 1 + 
\alpha^{(0)}_2 \sum_{\nu} \hat{p}_{\nu}^2 + 
\alpha^{(0)}_3 \sum_{\nu} \hat{p}_{\nu}^4 + 
\alpha^{(0)}_4 \sum_{\nu \neq \rho} \hat{p}_{\nu}^2 \hat{p}_{\rho}^2 + 
O(a^6). 
\end{equation}
For higher $i>0$, there are of course less terms relevant for a given order.
In general, all the possible covariant polynomial can be found through a character's
projection of the polynomial representation of the Hypercubic group onto the
defining (four dimensional) representation of the same group 
(see for instance \cite{FultonHarris} for a general reference).

To gain insight into Eq.~(\ref{hypTayl}), remember that in the free case the 
$\hat{\Sigma}_V^{(i)}$ correspond to the coefficient of $\hat{p}_{\mu}^{(i+1)}$
in the Taylor expansion of $2 \sin (\frac{\hat{p}_{\mu}}{2})$.
Eq.~(\ref{hypTayl}) is what we call a \emph{hypercubic-invariant Taylor expansion}.
The term in which we are interested is the leading term $\alpha^{(0)}_1$ in 
$\hat{\Sigma}_V^{(0)}$, since the other vanish in the continuum limit. In fact, the 
quark field RC $Z_q$ is defined as (see Eq.~(\ref{Zq}); in the following we assume the 
color average has already been done):
\[
Z_q(\mu a)  = \frac{
{\rm Tr_{spin}}( \sum_{\nu} \gamma_{\nu} p_{\nu}\hat{\Sigma}_V (\hat{p},\hat{m}_{cr},\beta^{-1}))
}{4 i p^2}\Big|_{p^2 = \mu^2}.
\]

In order to explain in details our fit procedure, let us define the auxiliary quantities:
\begin{equation}
\label{sigmak}
\sigma(k,\hat{p})= \frac{1}{M} \sum_{\mu \;: \;\hat{p}_{\mu}=\frac{2 a \pi}{L}\,k}
\frac{{\rm Tr_{spin}}(\gamma_{\mu} \hat{\Sigma}_V(\hat{p},\hat{m}_{cr},\beta^{-1})) 
}{\hat{p}_{\mu}} 
\end{equation}
($L = N a$ is the linear size of the the lattice). For a given momentum an average is taken 
over all the $M$ directions $\mu$ such that $\hat{p}_{\mu}=\frac{2 a \pi}{L} \, k$. 
For instance, consider the momentum $\hat{q} = (1,1,3,2) \; 2\pi/N$. 
In this case $\sigma(1,\hat{q})$ has two contributions ($M=2$): one from $\gamma_1$ 
and one from $\gamma_2$.  
Since also the $\sigma(k,.)$ are hypercubic-invariant, they can be averaged accordingly. 
Referring to the example above, consider $\hat{t} = (3,2,1,1) \; 2\pi/N$; simmetry requires that $\sigma(1,\hat{q}) = \sigma(1,\hat{t})$. In practice, they are averaged.
Notice that the functions $\sigma(k,.)$ 
are specific linear combinations of the $\hat{\Sigma}_V^{(i)}$. In practice we fit the 
data against the functions $\sigma(k,.)$. This is nothing but a fit of the constants
entering the parametrization of $\hat{\Sigma}_V^{(i)}$ in (\ref{hypTayl}).
If we include $O(a^4)$ terms (as in Eq.~(\ref{hypTayl})), we have to fit
7 unknown constants. To order $O(a^6)$ we have 14. 
We tried different orders up to  $O(a^6)$ and check the stability of the result.
The interesting term is the leading one. Other coefficients have to do with irrelevant effects. 

\begin{figure}[ht]
\begin{center} 
\includegraphics{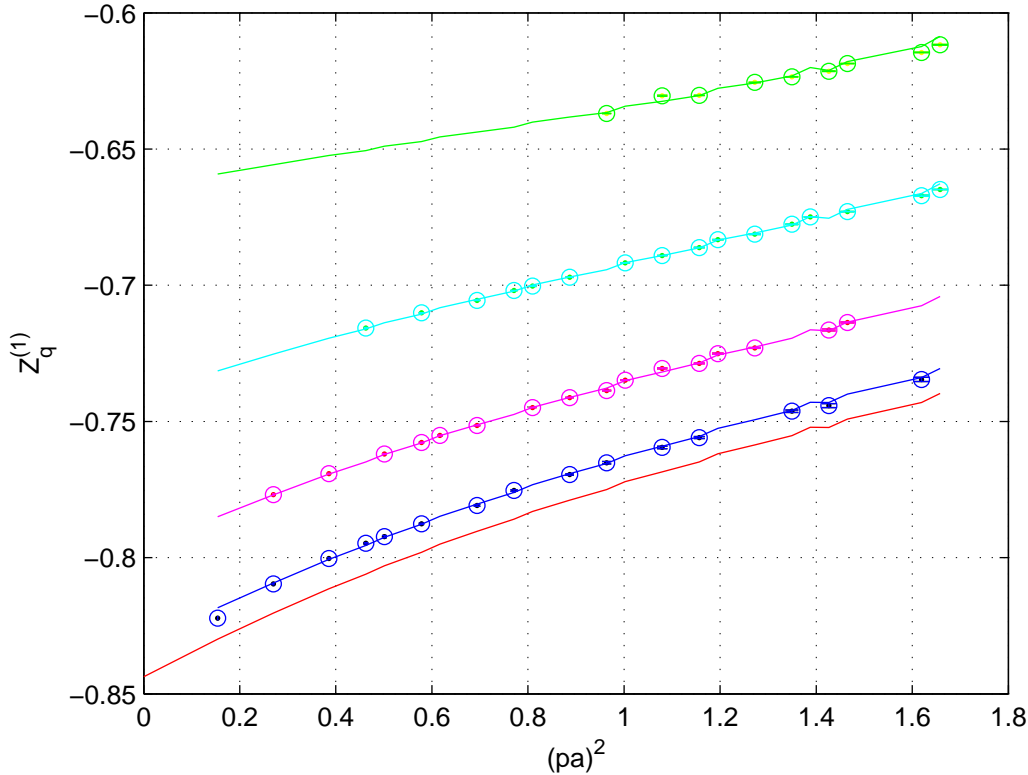}
\caption{\label{Fig.1}
The continuum-limit extrapolation of $Z_q^{(1)}$ (first loop of the quark field 
renormalization constant). We are interested in the intercept at $(pa)^2=0$, 
reached on the lowest line, which is the contribution 
$\hat{\Sigma}_V^{(0)}(\hat{p},\hat{m}_{cr},\beta^{-1})$ in Eq.~(\ref{hypTayl}).
The (blue, violet, azure-blue, green) curves represent the functions $\sigma(k,\hat{p})$ of 
Eq.~(\ref{sigmak}) for $k=1,2,3,4$. 
The red curve is $\hat{\Sigma}^{(0)}$, which is only meant to guide the eye to 
the intercept at $(pa)^2=0$.
The displayed fit include up to $O(a^6)$ terms. Stability has been checked with respect to 
different number of terms and intervals. 
}
\end{center}
\end{figure}

A nice illustration of the control that we have over our fits is provided by
Figure \ref{Fig.1}. There, we plot the functions $\sigma(k,\hat{p})$ up to $k=4$,
along with the curve $\hat{\Sigma}^{(0)}(\hat{p},\hat{m}_{cr},\beta^{-1}))$, 
whose intercept in $\hat{p}^2=0$ is the parameter we are looking for.
In Figure \ref{Fig.1} we choose to plot data versus $\hat{p}^2$, 
but one should keep in mind that this is not the only invariant under the 
hypercubic group entering
$\hat{\Sigma}_V^{(i)}(\hat{p},\hat{m}_{cr},\beta^{-1})$.
Our numerical result reproduces very well the analytical one.
%
%

\section{Results}

In the previous section we saw an example with no anomalous dimension.  
This is of course also the case for finite RC ($Z_V$ and $Z_A$) or finite 
ratios ($Z_P/Z_S$ and $Z_V/Z_A$). In the following we present our results for these quantities.
We computed at every order the relevant expectations values dictated by Eq.~(\ref{basicC}).
Finally we performed the amputation and the projection on the tree level structure. 
We could thus get the order-by-order expansions of the 
$O_{\Gamma}(pa)$'s in terms of which RC are defined. One-loop analytical results are well 
reproduced \cite{MartiZ}.

In the last subsection our results for the critical mass are presented.

%
\begin{figure}[hb]
  \begin{center} 
		\includegraphics[scale=0.55]{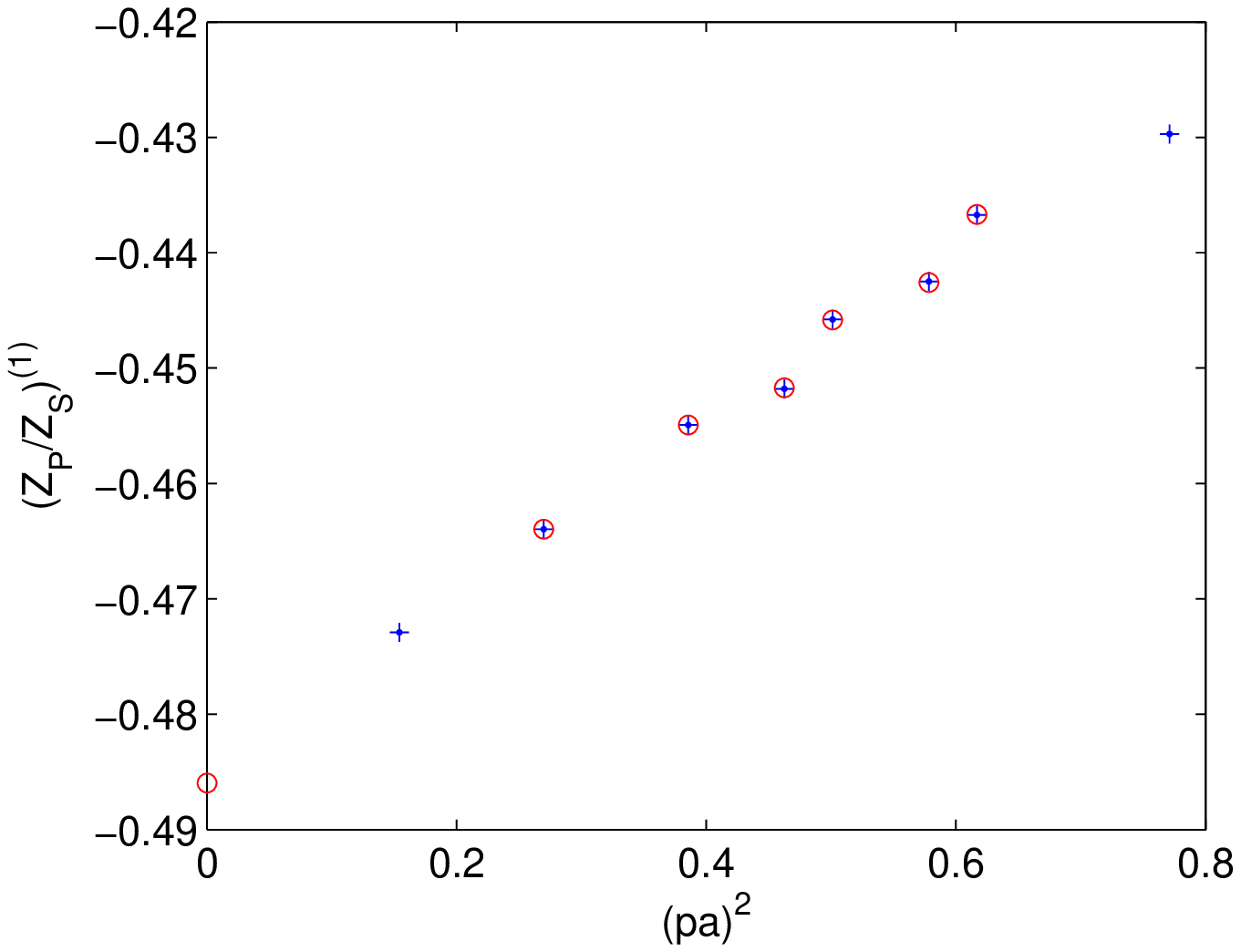}
		\includegraphics[scale=0.55]{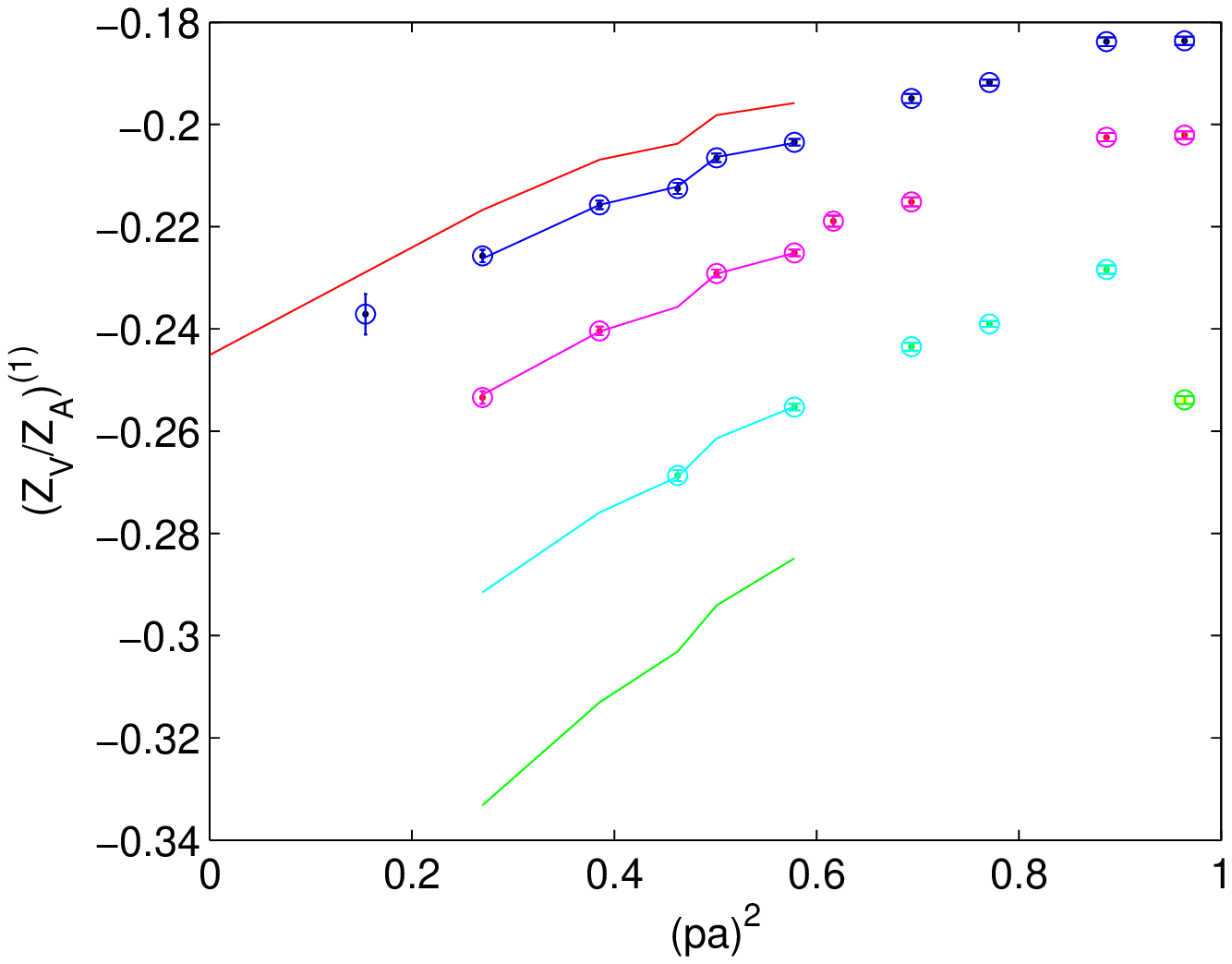}
    \caption{Computation to one loop of finite ratios of renormalization constants: $Z_P/Z_S$ (left) and $Z_A/Z_V$ (right).  Data points taken into account in these particular fits are enclosed in circles (left) or joined by solid lines (right; see caption of Fig.~1).}
   \label{Fig_PS_VA}
  \end{center}
\end{figure}%

\subsection{The finite ratios $Z_P/Z_S$ and $Z_V/Z_A$}\label{ratios}

The ratios $Z_P/Z_S$ and $Z_V/Z_A$ are safely computable at every order. This simply means to take (again, order by order) ratios of $O_{\Gamma}(pa)$ quantities. The quark field renormalization constant present in Eq.~(\ref{master}) drops out in the ratios, together with the divergence that affects $Z_P$ and $Z_S$ separately. In the end, one is left with the same situation we saw in the previous section: we simply have to perform  at every order \emph{hypercubic-invariant Taylor expansions} to get the continuum-limit coefficients of the expansions. 
One-loop examples are presented in Fig.~2. Fitting a scalar quantity like $Z_P/Z_S$ is actually easier (there is no direction singled out and consequently only one function, to be fitted as a polynomial in the hypercubic invariants).

We could perform many checks on our results. Finite-size effects are well under control, as checked by comparing results on $16^4$ and $32^4$ lattices in the quenched case. In the next section we will elaborate on computations for which this is not the case. We also stress that we can compute both $Z_A/Z_V$ and $Z_V/Z_A$; in the same way, we can compute both $Z_P/Z_S$ and $Z_S/Z_P$. Due to the order-by-order nature of the computation, this is not a tautology: different ratios come from different (although correlated) combinations of data. We checked that to a very good precision the series obtained are inverse of each other. 

Table~1 collects our results for different numbers of flavors. In the case $n_f=2$ four-loop results are available. As already pointed out, the fact that we were able to go one loop higher is due to our better knowledge of the three-loop critical mass in the $n_f=2$ case. Statistics in the cases $n_f=3,4$ is actually poorer. The fact that we could anyway go to three loops is a numerical accident: the signals for these ratios are actually very clean.

\begin{table}[hb]
\begin{center}
\begin{tabular}{|c|c|c|c|c|}
\hline
$Z_P/Z_S$ & & & & \\
\hline
\hline
$n_f$ &  $O(\beta^{-1})$  &   $O(\beta^{-2})$  &   $O(\beta^{-3})$  &   $O(\beta^{-4})$    \\
\hline
\hline
0   & - 0.487(1)  & - 1.50(1) & - 5.72(3) &  n.a.\\
\hline
2   &  - 0.487(1)  & - 1.46(1) & - 5.35(3) & - 21.6(3)  \\
\hline
3   & - 0.487(1)  & - 1.43(1) & - 5.13(3) &  n.a.\\
\hline
4   &  - 0.487(1)  & - 1.40(1) & - 4.86(3) &  n.a.\\
\hline
\hline
$Z_V/Z_A$  & & & &\\
\hline
\hline
$n_f$ &  $O(\beta^{-1})$  &   $O(\beta^{-2})$  &   $O(\beta^{-3})$  &   $O(\beta^{-4})$    \\
\hline
\hline
0   & - 0.244(1)  & - 0.780(5) & - 3.02(2) &  n.a.\\
\hline
2   &  - 0.244(1)  & - 0.759(5) & - 2.83(2) &  - 11.5(2) \\
\hline
3   & - 0.244(1)  & - 0.744(6) & - 2.72(2) &  n.a.\\
\hline
4   &  - 0.244(1)  & - 0.732(6) & - 2.57(2) &  n.a.\\
\hline
\end{tabular}
\vskip 0.5cm
\caption{The ratios $Z_P/Z_S$ and $Z_V/Z_A$ for various number of flavor $n_f$. Four-loop results are only available for $n_f=2$.}
\label{T_PS_VA}
\end{center}
\end{table}

Having results for various numbers of flavors one can proceed to fit the $n_f$-dependence. Since the polynomial dependence on $n_f$ of every order is fixed, this is another test for our results (see Fig.~3). We got
\[
	(Z_P/Z_S)^{(2)} =  - 1.50(1) + 0.0249(2)\, n_f  \;\;\;\;(Z_P/Z_S)^{(3)} =  - 5.72(3) + 0.151(5)\, n_f + 0.0159(5)\, n_f^2
\]
\[
	(Z_V/Z_A)^{(2)} =  - 0.780(5) + 0.0121(1)\, n_f  \;\;\;\;(Z_V/Z_A)^{(3)} =  - 3.02(2) + 0.073(2)\, n_f + 0.098(3)\, n_f^2
\]

%
\begin{figure}[t]
  \begin{center} 
		\includegraphics[scale=0.55]{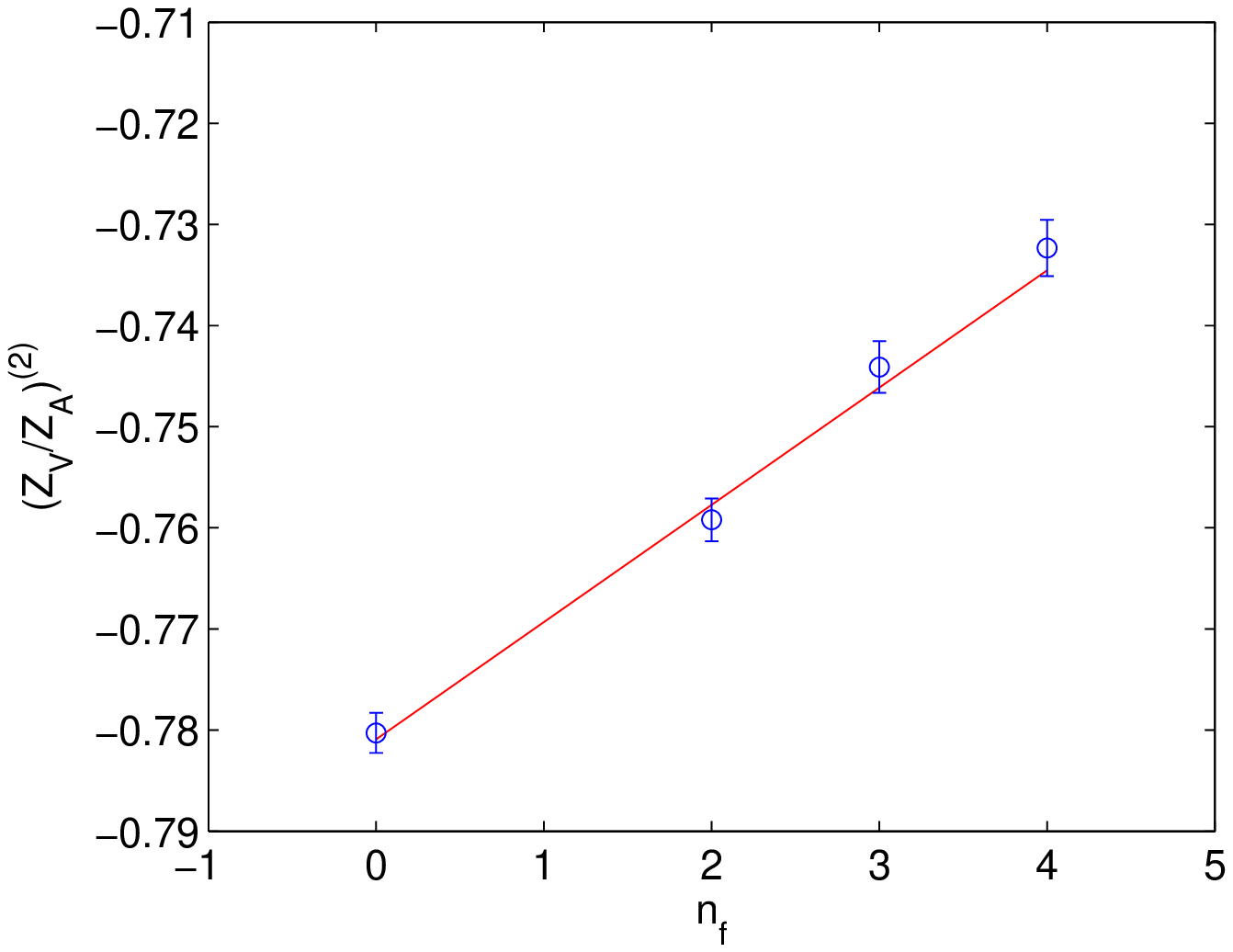}
		\includegraphics[scale=0.55]{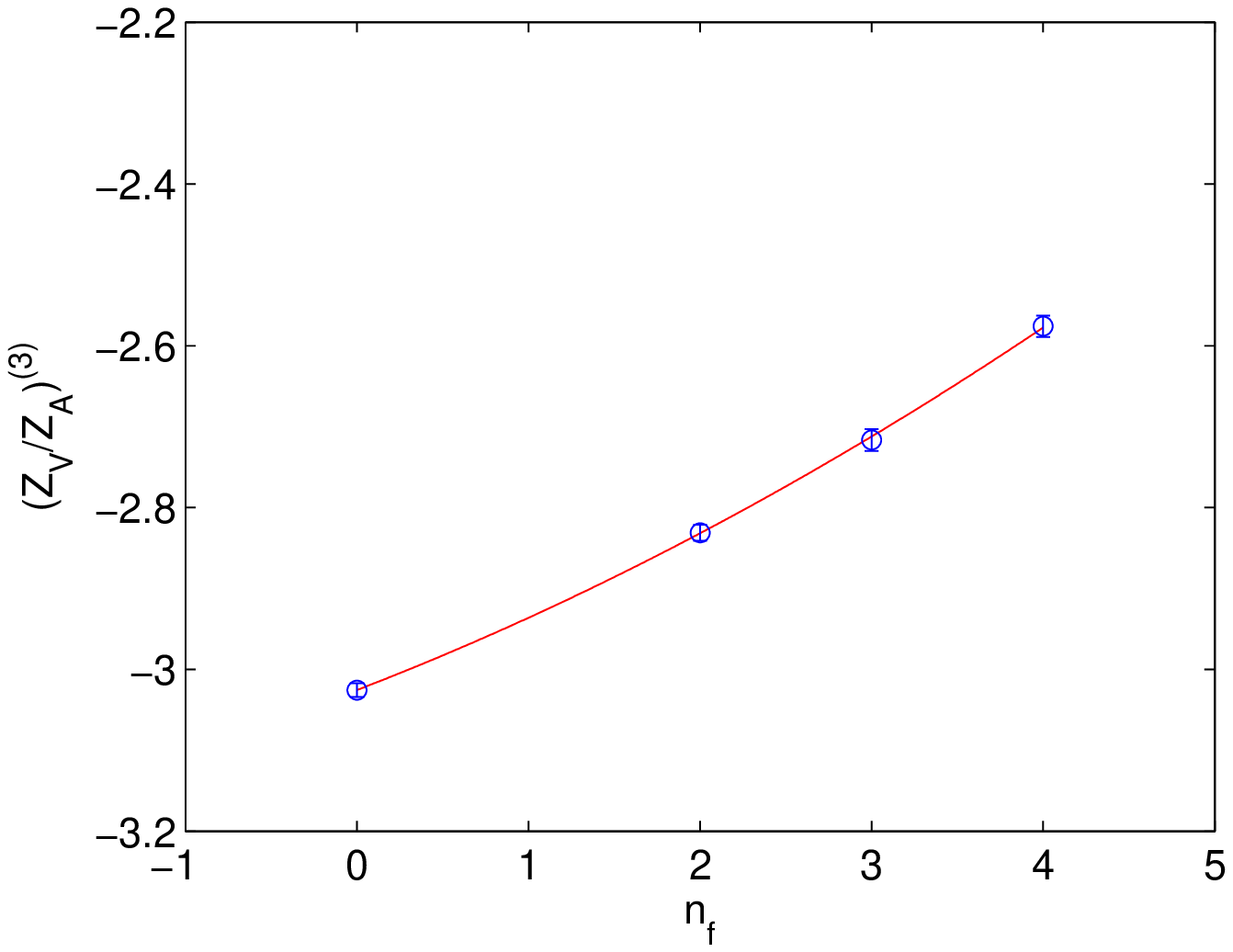}
    \caption{The $n_f$ dependence of the ratio $Z_V/Z_A$ at two (left, linear fit) and three (right, quadratic fit) loops.}
   \label{Fig_nf}
  \end{center}
\end{figure}%

Presented in this (more universal) way the precision of our results appears a bit poorer. As expected, results are dominated by quenched contributions. 

\subsection{$Z_V$ and $Z_A$}\label{A_V}

One loop examples of computations of $Z_V$ and $Z_A$ are plotted in Fig.~4. $Z_V$ and $Z_A$ are finite quantities by themselves. 
In our master formula Eq.~(\ref{master}) they are interlaced with log's coming from the quark field renormalization constant. The latter can be eliminated in two different ways. A first strategy is to cancel $Z_q$ directly from the measurements of the propagator. 
Another possibility is to take ratios with the conserved vector current: this is just what we did in the case of $Z_P/Z_S$ and $Z_V/Z_A$, this time having one of the $Z$'s equal to one. Both procedures return consistent results, which are summarized in Table~2, where we present results for $n_f=0,2$ (also in this case, four-loop results are available for $n_f=2$).

%
\begin{figure}[hb]
  \begin{center} 
		\includegraphics[scale=0.55]{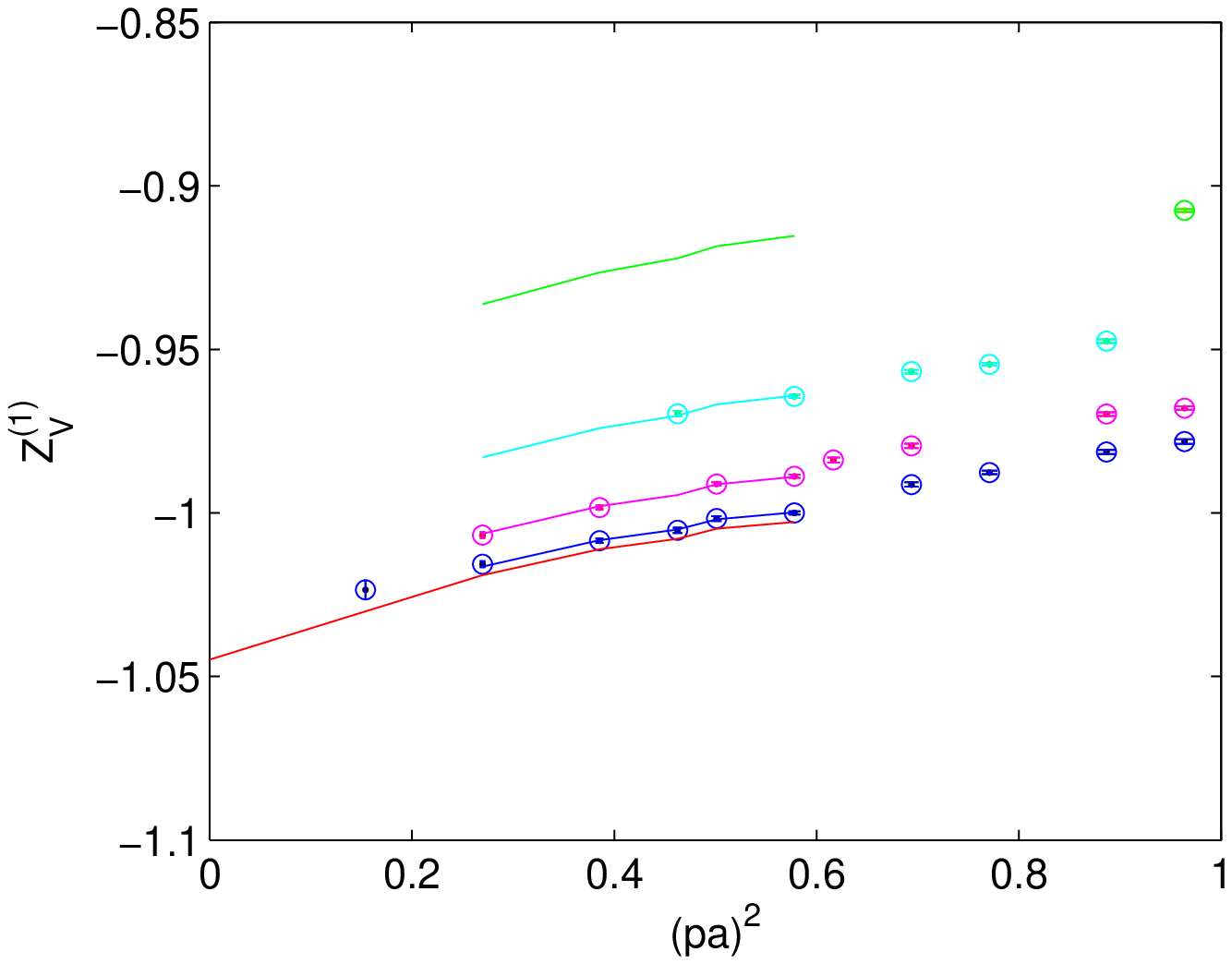}
		\includegraphics[scale=0.55]{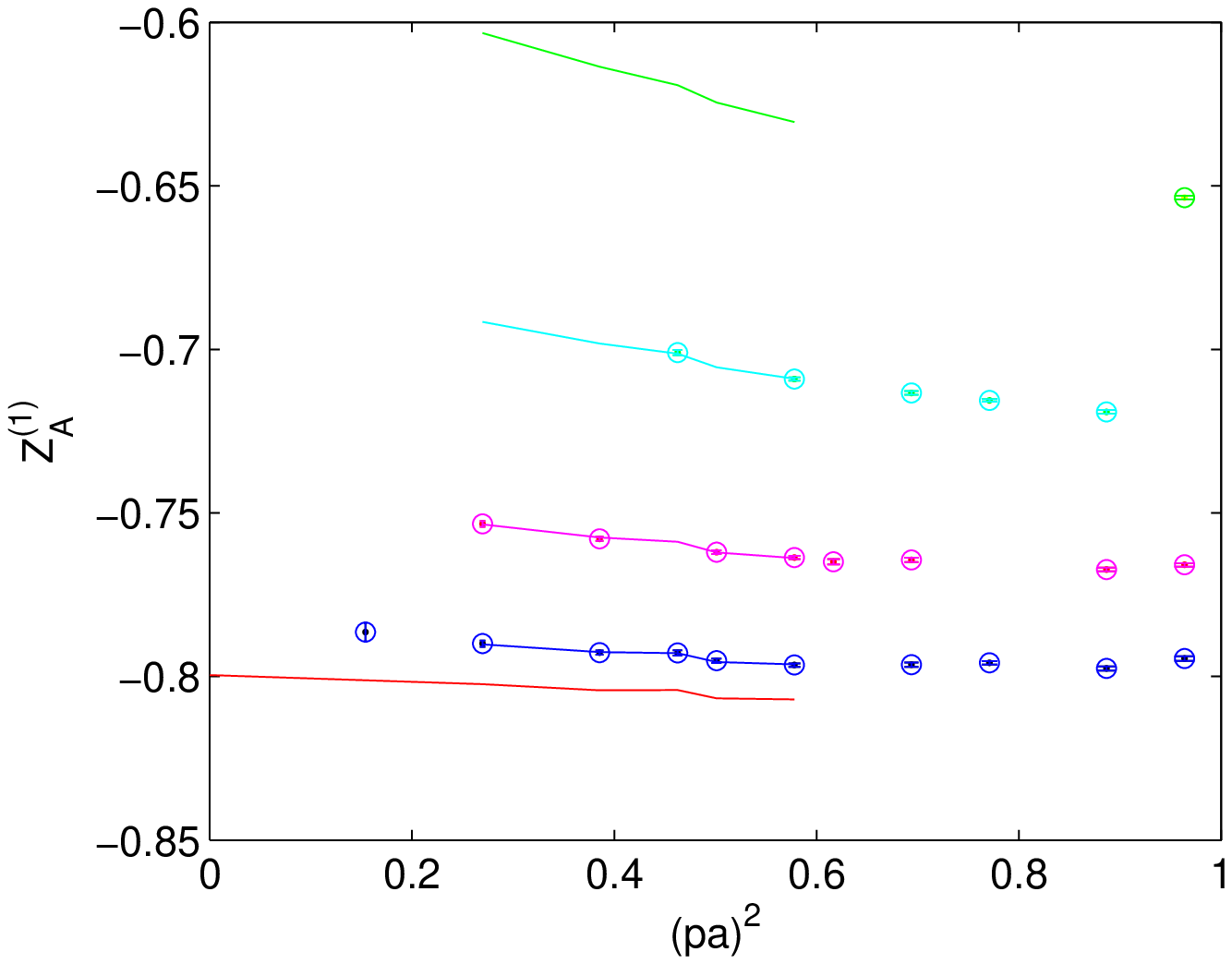}
    \caption{Computation to one loop of finite renormalization constants: $Z_V$ (left) and $Z_A$ (right). Same notations as in Fig.~2.}
   \label{Fig_A_V}
  \end{center}
\end{figure}%

\begin{table}[ht]
\begin{center}
\begin{tabular}{|c|c|c|c|c|}
\hline
$Z_V$ & & & & \\
\hline
\hline
$n_f$ &  $O(\beta^{-1})$  &   $O(\beta^{-2})$  &   $O(\beta^{-3})$   &   $O(\beta^{-4})$ \\
\hline
0   & - 1.044(2)  & - 1.98(3) & - 6.10(8) & n.a.\\
\hline
2   &  - 1.044(2)  & - 1.88(3) & - 5.42(8)  &  - 17.0(9)  \\
\hline
\hline
$Z_A$  & & & &\\
\hline
\hline
$n_f$ &  $O(\beta^{-1})$  &   $O(\beta^{-2})$  &   $O(\beta^{-3})$   &   $O(\beta^{-4})$ \\
\hline
0   & - 0.800(2)  & - 1.39(3) & - 4.04(4) & n.a.\\
\hline
2   &  - 0.800(2)  & - 1.31(3) & - 3.50(8)  &  - 9.8(6) \\
\hline
\end{tabular}
\vskip 0.5cm
\caption{The finite renormalization constants $Z_V$ and $Z_A$ for $n_f=0,2$.}
\label{T_V_A}
\end{center}
\end{table}

We have just discussed the two different approaches we used to compute $Z_V$ and $Z_A$. In the previous subsection we presented results for the ratio $Z_V/Z_A$, which can of course as well be computed from the computation of $Z_V$ and $Z_A$. One can verify that all these measurements are very well consistent. Still, they are controlled by different numerical noise, so that (for example) a direct computation of the ratio $Z_V/Z_A$ is viable for all the $n_f$ we took into account, while this is not the case for $Z_V$ and $Z_A$ separately (as already stated, statistics for $n_f=3,4$ is poorer). In the end, all these procedures differ from each other for different ways of fitting irrelevant contributions. Getting rid of irrelevant contributions to single out continuum-limit results is a key issue in our approach, and so consistency between all these computations is a good test for reliability of our results. 

\subsection{A by-product: the critical mass}

Analytical computations of the critical mass are available up to two loops \cite{PanaPelo}. A three-loop computation in the $n_f=2$ case was reported by our group in \cite{NSPTfull}. Here we present three-loops result for other $n_f$ and add a four loop result for $n_f=2$. Results are collected in Table~3. They were obtained from the defining formula of Eq.~(\ref{Mc}) by fitting irrelevant contributions to $\hat{\Sigma}_c(\hat{p},\hat{m}_{cr},\beta^{-1})$. Also in this case there was no log coming from an anomalous dimension: in this case there is a power divergence, in force of which a perturbative result is not to be taken as an accurate one. It is nevertheless valuable indeed to maintain massless our fermions, \emph{i.e.} as a counterterm.

\begin{table}[h]
\begin{center}
\begin{tabular}{|c|c|c|c|c|}
\hline
$m_{cr}$ &  $n_f=0$  &   $n_f=2$  &   $n_f=3$   &   $n_f=4$ \\
\hline
$O(\beta^{-3})$   & - 13.11(6)  & - 11.78(5) & - 11.02(9) & - 10.24(9).\\         
\hline
$O(\beta^{-4})$   &  n.a  & - 39.6(4) & n.a  &  n.a  \\
\hline
\end{tabular}
\vskip 0.5cm
\caption{Three-loop critical mass for various $n_f$; a four-loop result is available for $n_f=2$.}
\label{T_Mc}
\end{center}
\end{table}

\noindent Also in this case, one can fit a generic $n_f$ result
\[
	m_{cr}^{(3)} =  - 13.11(6) + 0.62(5)\, n_f + 0.024(9)\, n_f^2 \,.
\]

\section{Dealing with anomalous dimensions}
\label{sec:anomdim}

We anticipated that dealing with anomalous dimensions requires some extra care. In order to get some insight, we discuss a first example in which an anomalous dimension comes into place, \emph{i.e.} the one-loop computation of $Z_S$. In this case our master formula Eq.~(\ref{master}) reads
\[
	\left( 1 - {{z_q^{(1)}}\over{\beta}} + \ldots \right) \left( 1 + {{z_s^{(1)}- \gamma_s^{(1)} \log(\hat{p}^2)}\over{\beta}} + \ldots \right) \left( 1 + {{O_s^{(1)}(\hat{p}^2)}\over{\beta}} + \ldots \right) \Big|_{p^2 = \mu^2} \, = \, 1	
\]
in which we explicitly wrote both the constant and the logarithmic contributions to renormalization constants (the only log comes in this case from $Z_S$ since one-loop quark-field anomalous dimension is zero in Landau gauge). $O_s^{(1)}(\hat{p}^2)$ is what is actually numerically measured. At one-loop order we can solve the previous relation to
\begin{equation}\label{Zs}
	z_q^{(1)} - z_s^{(1)} =   O_s^{(1)}(\hat{p}^2) - \gamma_s^{(1)} \log(\hat{p}^2) \,.
\end{equation}
The message from Eq.~(\ref{Zs}) is simple: we will first subtract the logarithmic contribution and then proceed to our \emph{hypercubic-invariant Taylor expansion}. This is plotted in Fig.~5: upper data points are $O_s^{(1)}(\hat{p}^2)$, lower data points are the subtracted ones. We can see on the left of Fig.~5 that by going through this procedure we miss the analytical result. Notice that it looks like we were \emph{subtracting too much}. To be definite, the subtracted data points bend quite a lot in the IR region. In the end, this does not come as a surprise: \emph{RI'-MOM} is an infinite-volume scheme, but we are necessarily dealing with finite $N$ (number of lattice points) computations. 
Since for $n_f=0$ we have both $32^4$ and $16^4$ data, we are in a position to verify whether this is the real issue.

%
\begin{figure}[hb]
  \begin{center} 
		\includegraphics[scale=0.55]{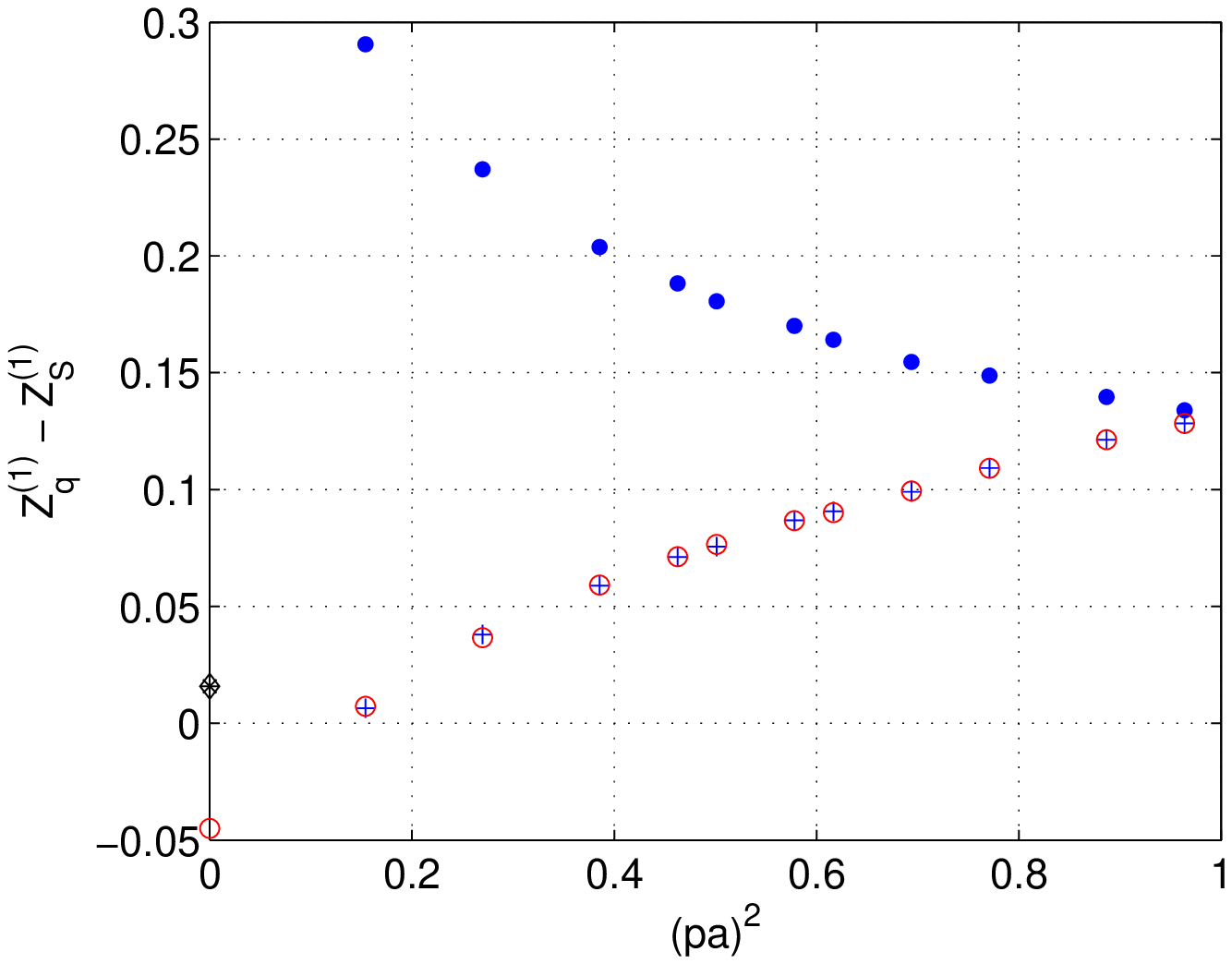}
		\includegraphics[scale=0.55]{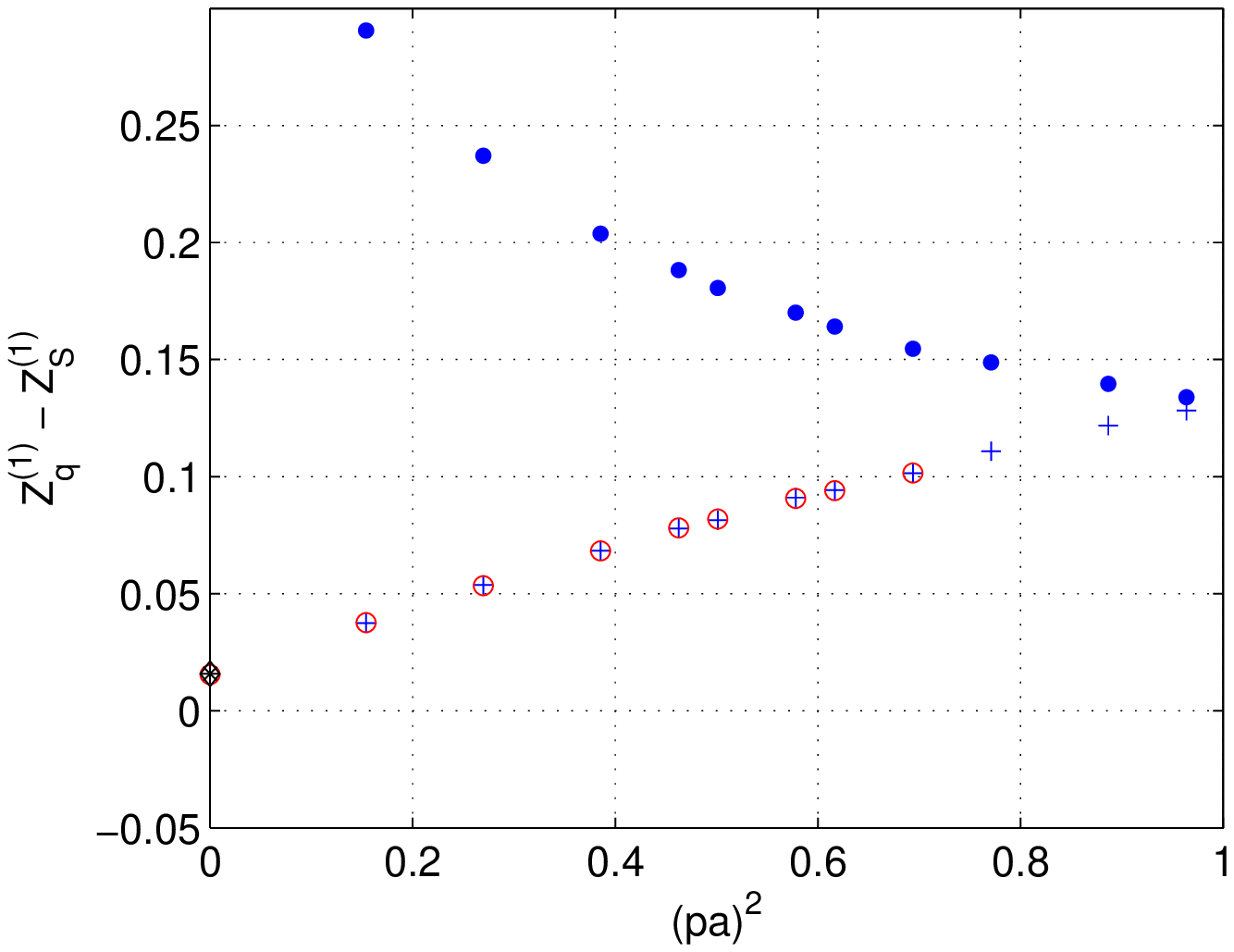}
    \caption{Computation of one loop renormalization constant for the scalar current. With respect to Eq.~(\ref{Zs}), upper points are the unsubtracted $O_s^{(1)}(\hat{p}^2)$, while lower (circled crosses) stand for the subtracted $O_s^{(1)}(\hat{p}^2) - \gamma_s^{(1)} \log(\hat{p}^2)$. Analytic result is marked with a darker symbol. On the left: no correction for finite volume. On the right: finite-volume \emph{tamed-log} taken into account.}
   \label{Fig_tamedLOG}
  \end{center}
\end{figure}%

%
\begin{figure}[t]
  \begin{center} 
		\includegraphics[scale=1]{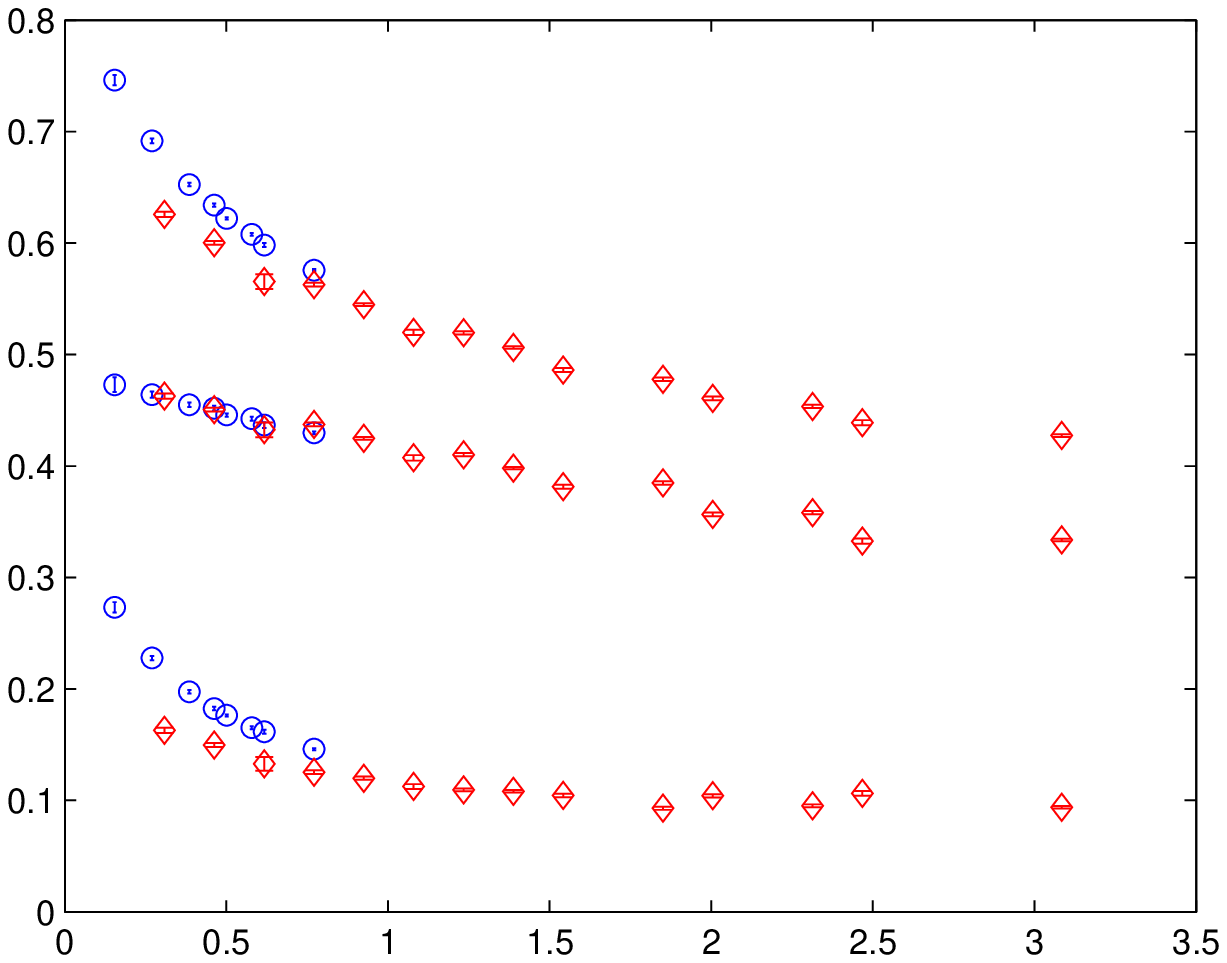}
    \caption{Computations of $O_p^{(1)}(\hat{p}^2)$ (the equivalent of Eq.~(\ref{Zs}) for the pseudoscalar current) (top) 
    and $O_s^{(1)}(\hat{p}^2)$ (bottom) on $32^4$ (circles) and $16^4$ (diamonds). In the middle the ratio $\frac{O_s^{(1)}(\hat{p}^2)}{O_p^{(1)}(\hat{p}^2)}$, which appears safe with respect to finite-size effects.}
   \label{sinossi_16_32_SP1L}
  \end{center}
\end{figure}%

Fig.~6 displays our results for $O_p^{(1)}(\hat{p}^2)$ (the equivalent of Eq.~(\ref{Zs}) for the pseudoscalar current), $O_s^{(1)}(\hat{p}^2)$ and of the ratio $\frac{O_s^{(1)}(\hat{p}^2)}{O_p^{(1)}(\hat{p}^2)}$ on the two different volumes. While the ratio (in the middle of the figure) is safe (we have already made this point in the previous Section), quite remarkable finite-size effects are manifest for the $O_i^{(1)}(\hat{p}^2)$. It is obvious that by performing the subtraction of Eq.~(\ref{Zs}) on the $16^4$ data points one misses the analytical result even more than in the left of Fig.~5. The picture stays much the same at higher loops.

An important caveat is in order. We have already made the point that our regularization of zero modes prescribes to remove the degrees of freedom associated to them. This is a legitimate procedure in the $N \rightarrow \infty$ limit, which in turn means that we can have better and better approximations of infinite volume results, but we can not aim at having consistent perturbative expansions at finite physical volume. 
Our aim is to single out the $N=\infty$ behavior, but this requires to confront finite-$N$ corrections, 
which we expect to be sizable in particular in the IR region.

One can define $L = N a$. Let us now write down for the quantity at hand the momentum sum $I(p,a,L)$ encoding the lattice Feynman diagram of the conventional Lattice Perturbation Theory, with the same 
\emph{ad hoc} regularization of zero modes 
(zero momentum removed from the sum). Dimensional analysis suggests the presence of 
$pL = \hat{p}N$ effects (this relation holds for every value of $a$). In the spirit of the 
famous work \cite{Kawai} one can now split a (logarithmically divergent) Feynman diagram as in
\begin{equation}
	I(p,a,L) = I(0,a,L) + \left( I(p,a,L)- I(0,a,L) \right) \equiv I(0,a,L) + J(p,a,L).
\end{equation}
We can now manipulate the momentum sums. The divergence is logarithmic so that by subtracting $I(0,a,L)$ we make $J(p,a,L)$ UV finite. Therefore it can be computed (with the same \emph{ad hoc} regularization of zero modes) in the $a \rightarrow 0$ limit. Although this does not define a finite volume perturbative 
computation, it is a legitimate manipulation of the sum. In general, it will be now IR divergent, but this divergence (which is anyway regularized by finite L) will be canceled by contributions coming from $I(0,a,L)$, \emph{i.e.}
\begin{eqnarray}
	I(0,a,L) & = & c_1 + \gamma \, log(a/L) + H(a/L) \\ \nonumber
	J(p,a,L) & = & c_2 + \gamma \, log(pL) + G(pa,a/L,pL)\,.
\end{eqnarray}
We point out that $I(0,a,L)$ can not contain $pL$ effects: these should be looked for in $J(p,a,L)$. Therefore one can look for $pL = \hat{p}N$ effects in $G(pa,a/L,pL) \rightarrow \tilde{G}(pL)$. In order to obtain this quantity, we just computed the relevant graph in the formal continuum limit of our sum $J(p,a,L)$ ($a \rightarrow 0$ with $L = N a$ fixed), with the same \emph{ad hoc} regularization of zero modes. We call this contribution a \emph{tamed-log}, 
since it is supposed to resemble the expected log, but with $pL = \hat{p}N$ effects on top of it. 
We find that this function indeed approaches a log for $p>>1$. Fig.~5 displays our results once one subtracts this \emph{tamed-log}. As a matter of fact, if one stays away from deep IR the subtracted data points on the left and on the right of Fig.~5 are much the same. We stress that we are not saying that the finite $N$ effects we have just elaborated on are the only ones. By inspection, they appear to be the relevant ones, as it confirmed by the fact that $N=32$ and $N=16$ now return the same results.

The situation is more complicate at higher loops. We will devote to it a separate paper \cite{ourNEXT}, in which we will explicitly gain informations from different lattice sizes.

\section{Resumming the series}

We now go back to the expansions of subsection \ref{ratios} and \ref{A_V} and try to resum them to obtain the finite RC's.  
Giving results and errors on top of them requires the estimation of truncation errors. We will in the following adopt the strategy of BPT. We stress from the very beginning that our real goal is the estimation of convergence properties of the series. It is only in force of the sufficiently high order of the expansions that one can hope to really gain insight. One should nevertheless be ready to accept that every statement on convergence will be decided on a strict case-by-case policy.

The different coupling constants we will use are all obtained in terms of the basic plaquette $P$. Let us define
\begin{equation}\label{BPTcouplings}
	x_0 = \beta^{-1} \;\;\;\; x_1 \equiv \frac{\beta^{-1}}{\sqrt{P}}  
\;\;\;\; x_2 \equiv - \frac{1}{2} \log(P)  \;\;\;\; x_3 \equiv \frac{\beta^{-1}}{P} \,.
\end{equation}
$x_2$ and $x_3$ are quite popular as boosted couplings. The reason why we also define $x_1$ will be clear in a moment. Obtaining the expansions in $x_i$ once the expansions in $x_0$ are known is a textbook exercise, given the definitions in Eq.~(\ref{BPTcouplings}). One needs the expansion of the plaquette: analytical results \cite{TomeuSandraHaris} are only known to a given order, but our simulations always provide also the expansion of $P$.

We resum the series at $\beta=5.8, n_f=2$. This makes possible a comparison with the non-perturbative results of \cite{CeciVitt}. 

Fig.~7 displays the resummation of $Z_P/Z_S$ and $Z_S/Z_P$ in the four different couplings. One can inspect from the very beginning the impact of a basic property of BPT which is often underestimated: all the couplings are equal at tree level, which means that all the expansion are equal at leading order. One loop BPT amounts to sitting on a straight line, whose slope is dictated by the one-loop coefficient. Only at higher loops we can gain some insight on convergence properties. There is actually a variety of convergence patterns (taking also $x_1$ into account is helpful with this respect).

In particular, one can check the following:
\begin{itemize}
	\item Within a fixed definition of the coupling, convergence is of course better and better as the order increases. As common wisdom suggests, convergence in the bare coupling is not so brilliant and in general quite different convergence patterns are manifest; they appear quite satisfactory for $x_2$ and $x_3$. In particular, for the case of $x_2$ expansion we plot in Fig.~8 the deviations $\Delta^{(n)}$, defined as the differences between resummation at order $n$ and resummation at order $n-1$ . The good scaling should not be taken too seriously (this is largely a numerical accident). Still, this is signaling a reasonable convergence pattern.
	\item As the order increases, expansions in different couplings get closer to each other, as expected; in particular expansions in $x_2$ and $x_3$ are quite close to each other.  
	\item The resummed results for $Z_P/Z_S$ and $Z_S/Z_P$ in the $x_2$ and $x_3$ couplings are the inverse of each other to a reasonable accuracy. This is also a good indication.
\end{itemize}

%
\begin{figure}[ht]
  \begin{center} 
		\includegraphics[scale=0.45]{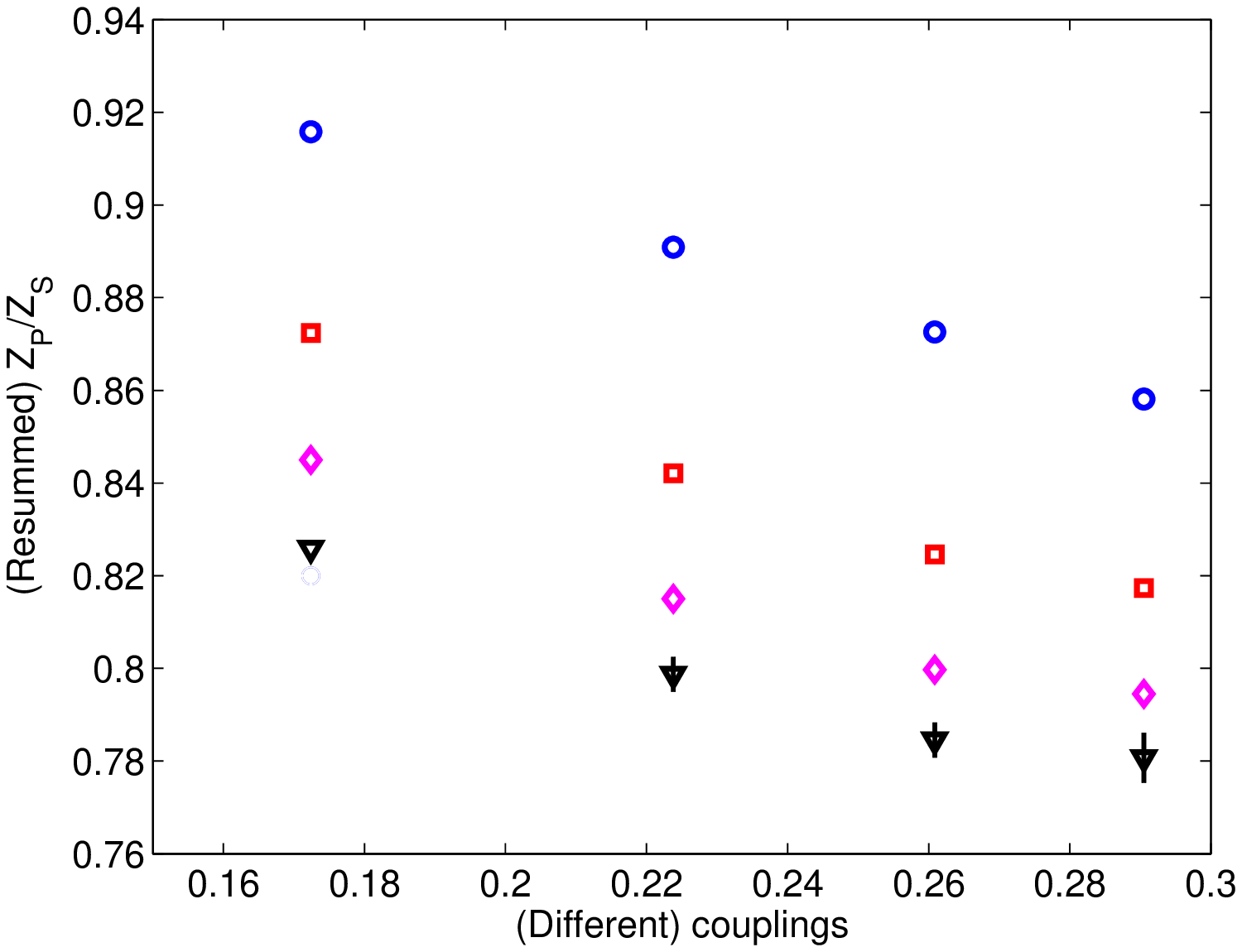}
		\includegraphics[scale=0.45]{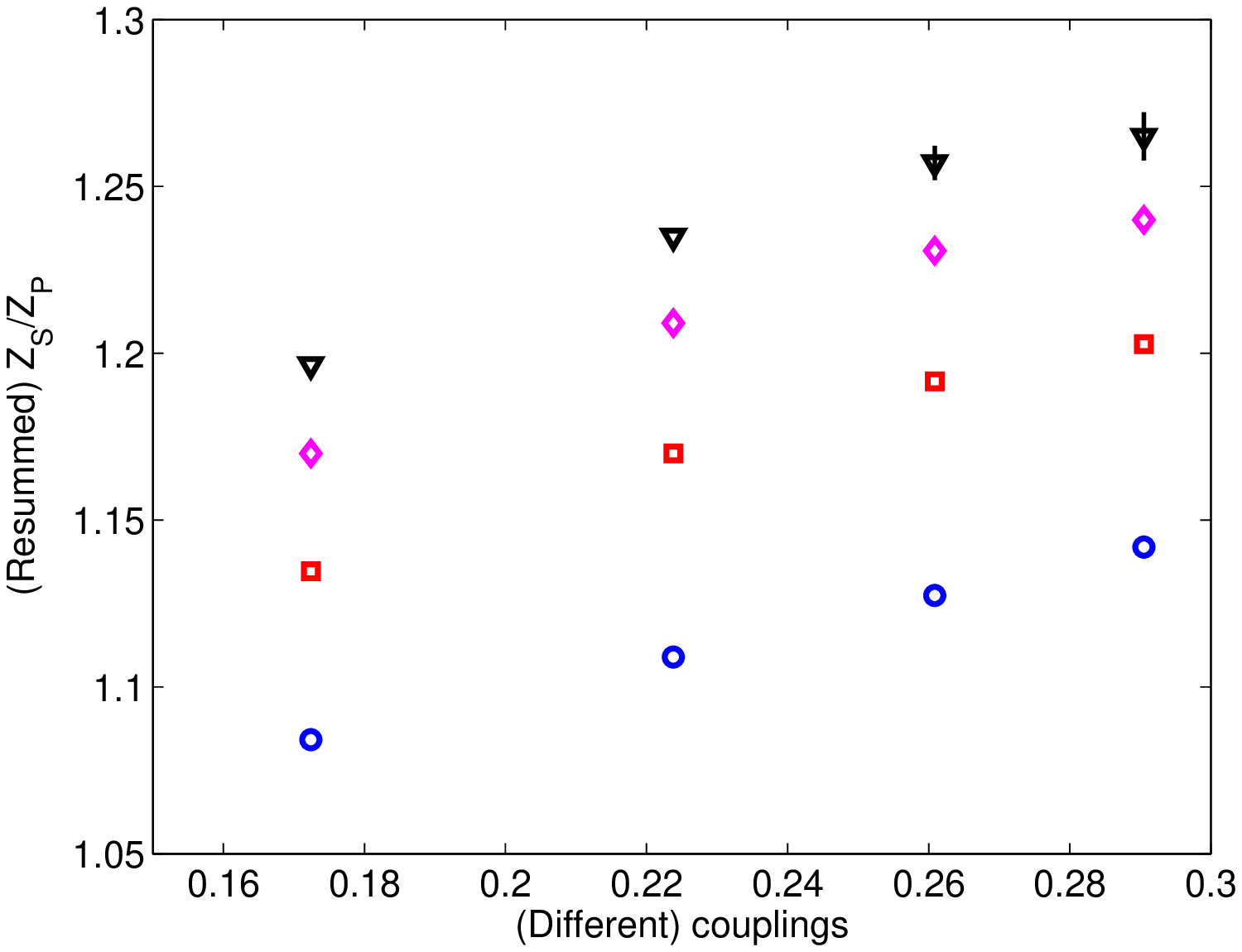}
    \caption{Resummations of $Z_P/Z_S$ (left) and $Z_S/Z_P$ (right) for $n_f=2$ at $\beta=5.8$ to one (circles), two (squares), three (diamonds) and four (triangles) loops (the last is the only one which has a sizable error). We show resummations for different couplings: on the $x$-axis, the (different) values of the different couplings. From the left: $x_0$, $x_1$, $x_2$, $x_3$ ($x_0$ is $\beta^{-1}$, see text for the definitions of the other couplings).}
   \label{Fig_resumPS}
  \end{center}
\end{figure}%

%
\begin{figure}[ht]
  \begin{center} 
		\includegraphics[scale=0.6]{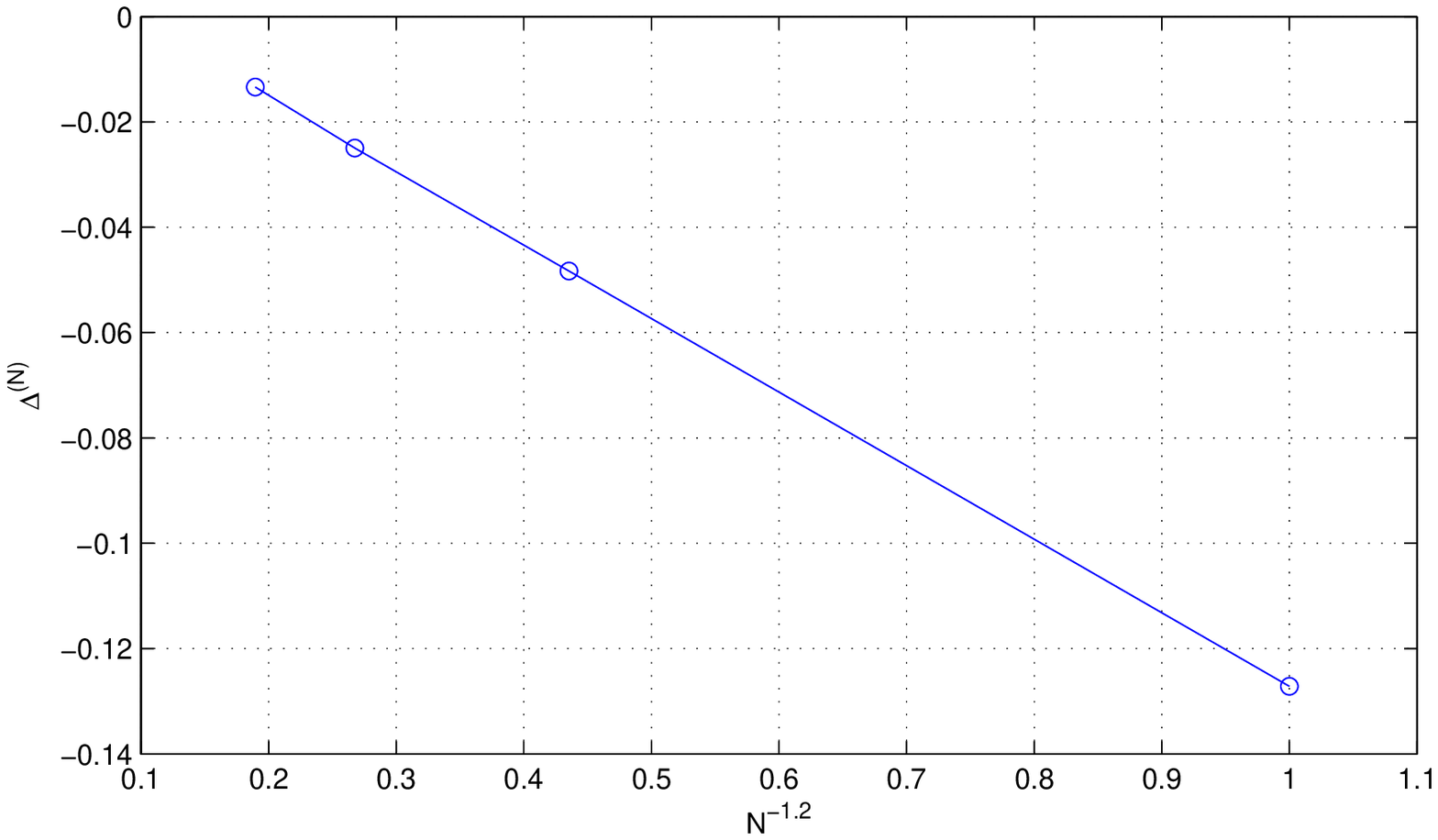}
    \caption{The scaling of deviations of different order truncations for the quantity $Z_P/Z_S$ for the $x_2$ coupling.}
   \label{Fig_scal}
  \end{center}
\end{figure}%
 
Convergence properties of the expansions in the $x_2$ and $x_3$ couplings are good enough to extract a result. We notice that if one adds to the result at a given order the deviation from the immediately lower order, one always ends up at the same result (as a matter of fact a popular way to pin down a truncation error is just taken from deviations which we previously called $\Delta_n$). We thus quote $Z_P/Z_S=0.77(1)$.

We have already made the point that to assess convergence properties one should adopt a case by case strategy. This can be clearly seen when we proceed to resum $Z_A$ and $Z_V$. Before doing that, we give a trivial example of what a blind application of the idea of BPT can result in. In Fig.~9 we \emph{exaggerate} the boosting of the coupling, by taking into account other coupling $x_\alpha\equiv \frac{\beta^{-1}}{P^\alpha}$ ($\alpha>1$). As one can see, convergence properties are completely jeopardized. 

%
\begin{figure}[hb]
  \begin{center} 
		\includegraphics[scale=0.5]{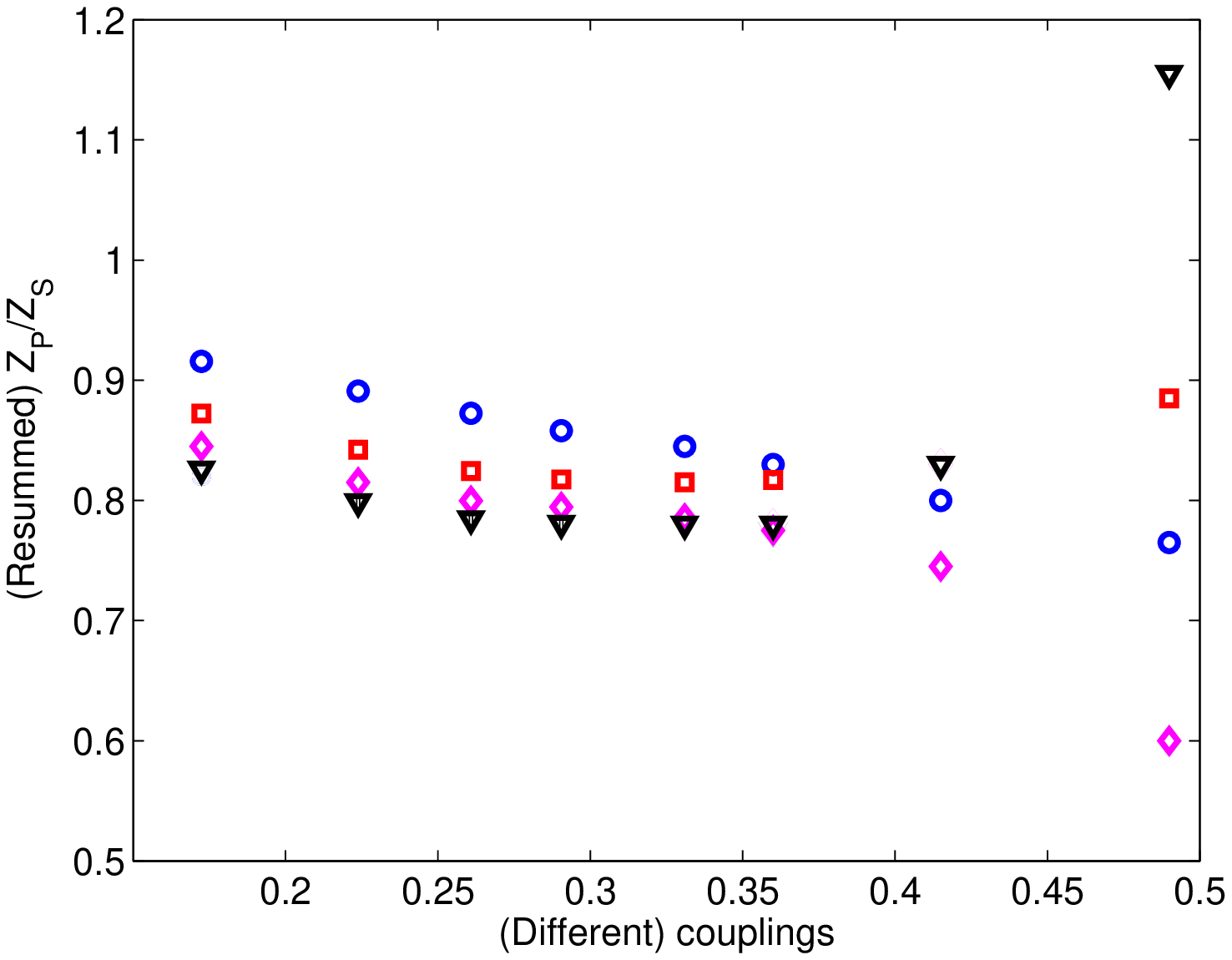}
    \caption{Resummations of $Z_P/Z_S$ (left) for $n_f=2$ at $\beta=5.8$. The same as in Fig.~7, but this time exaggerating the boosting of the couplings.}
   \label{Fig_BPTooMuch}
  \end{center}
\end{figure}%

%
\begin{figure}[t]
  \begin{center} 
		\includegraphics[scale=0.55]{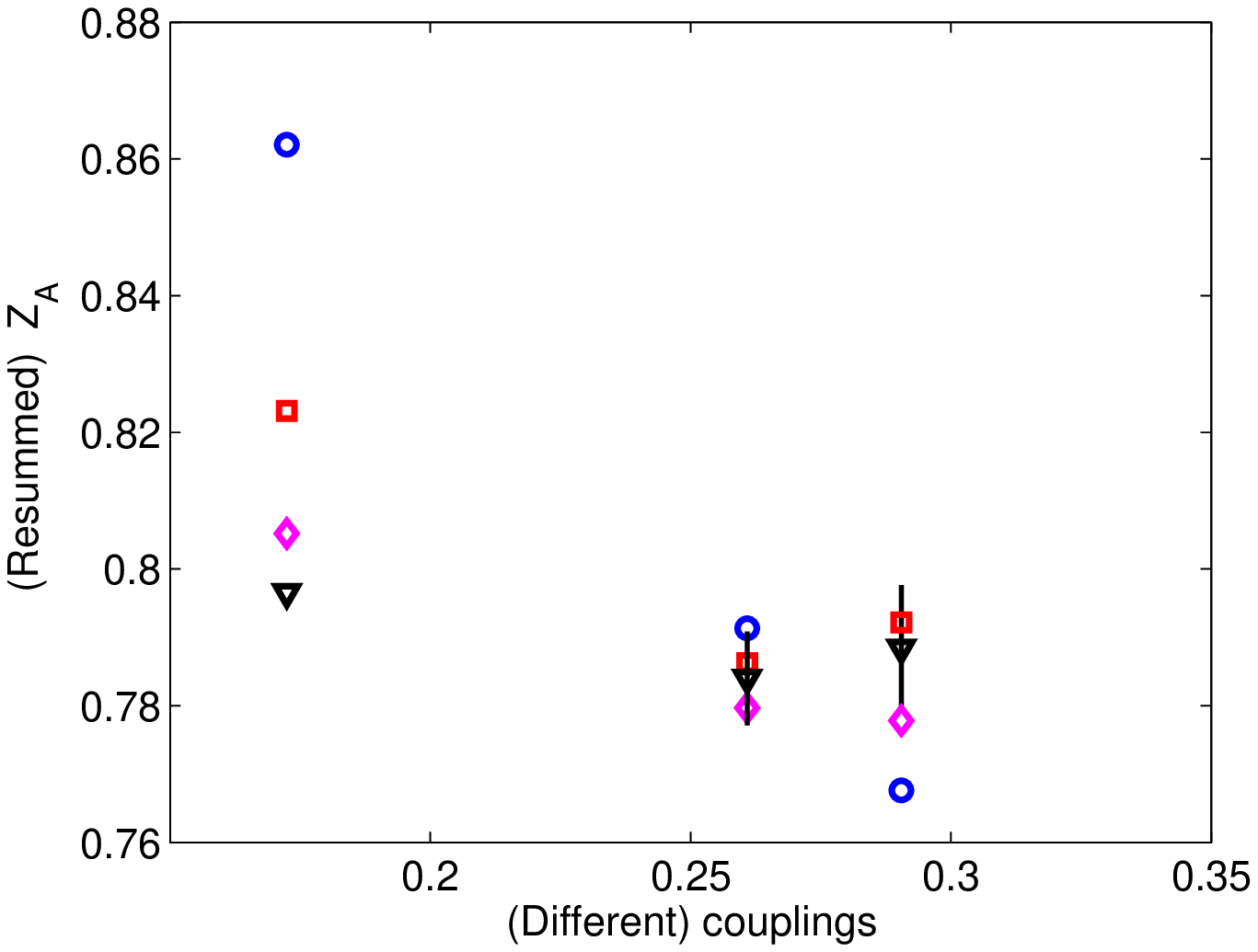}
		\includegraphics[scale=0.55]{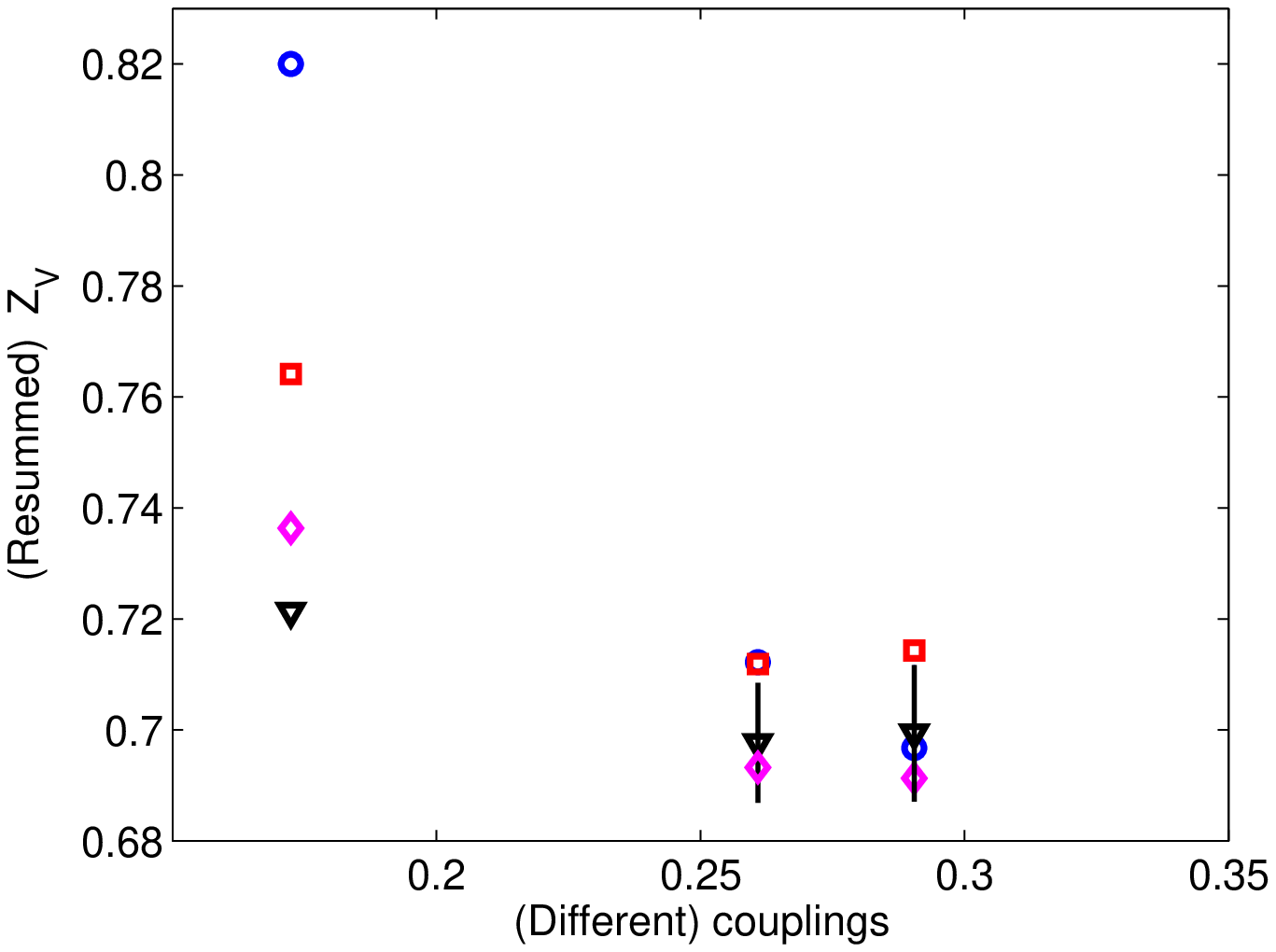}
    \caption{Resummations of $Z_A$ (left) and $Z_V$ (right) for $n_f=2$ at $\beta=5.8$ to one (circles), two (squares), three (diamonds) and four (triangles) loops (the last is the only one which has a sizable error). We show resummations for different couplings: on the $x$-axis, the (different) values of the different couplings. From the left: $x_0$, $x_2$, $x_3$ (see text for the definitions of the couplings).}
   \label{Fig_resum}
  \end{center}
\end{figure}%

In Fig.~10 we plot the resummation of $Z_V$ and $Z_A$ (again, at $\beta=5.8,n_f=2$). As one can see, this time convergence properties of the expansion in the bare coupling are not so bad. Consequently, one is already at risk of overshooting at one loop BPT and the expansions in $x_2$ and $x_3$ oscillate. Our final estimates are $Z_A=0.79(1)$ and $Z_V=0.70(1)$. 

Our resummed results are quite consistent with \cite{CeciVitt}. A bigger deviation is seen on the values of $Z_A$ and $Z_V$. To our understanding this could be mainly imputed to the indetermination coming from the chiral extrapolation.

\section{Conclusions}

We presented high-order computation of renormalization constants for Lattice QCD. Finite-size effects are well under control for the quantities we considered. There is no extrapolation involved in staying at the chiral limit in which renormalization conditions are imposed. The continuum limit extraction is achieved in a clean way. Truncation errors can be well assessed by a judicious use of BPT. Thus, the main message of this paper is that high precision perturbative computations of lattice QCD renormalization constants are feasible and should not be regarded necessarily as a second choice.

Further work will follow, both to complete the job for logarithmically divergent quantities and to take into account different actions (in particular different fermionic regularizations). This is not expected to imply any change in strategy and the implementation is mainly a matter of programming. In particular, work has already started to extend results to Clover fermions and to other gauge actions.

\section*{Acknowledgments}
\par\noindent
We warmly thank Andrea Mantovi for having collaborated with us at an early stage of this project. 
We are very grateful to V. Lubicz and C. Tarantino for many stimulating discussions and for sharing with us their data. 
We also acknowledge interesting discussions with S. Capitani. 
F.D.R., V.M. and C.T. acknowledge support from both Italian MURST 
under contract 2001021158 and from I.N.F.N. under {\sl i.s. MI11}. 
L.S. has been partially supported by DFG through the Sonderforschungsbereich
'Computational Particle Physics' (SFB/TR 9).


\end{document}